\documentstyle[psfig]{mn2e}
\newcommand{\etal}{{et al.~}}

\newcommand{\kms}{\>{\rm km}\,{\rm s}^{-1}}

\newcommand{\beq}{\begin{equation}}
\newcommand{\eeq}{\end{equation}}
\newcommand{\kpch}{\>{h^{-1}{\rm kpc}}}
\newcommand{\mpch}{\>h^{-1}{\rm {Mpc}}}

\newcommand{\absmag}{M_{^{0.1}r}}
\newcommand{\mpci}{\ $h$\ Mpc$^{-1}$\ }

\newcommand{\apj}{ApJ}

\newcommand{\aj}{AJ}
\newcommand{\mnras}{MNRAS}

\newdimen\hssize
\newdimen\hdsize 
\def\beq{\begin{equation}}
\def\eeq{\end{equation}}
\def\bey{\begin{eqnarray}}
\def\eey{\end{eqnarray}}

\def\kms{\,{\rm {km\, s^{-1}}}}

\def\gs{\mathrel{\raise1.16pt\hbox{$>$}\kern-7.0pt
\lower3.06pt\hbox{{$\scriptstyle \sim$}}}}
\def\ls{\mathrel{\raise1.16pt\hbox{$<$}\kern-7.0pt
\lower3.06pt\hbox{{$\scriptstyle \sim$}}}}
\def\gtsima{$\; \buildrel > \over \sim \;$}
\def\ltsima{$\; \buildrel < \over \sim \;$}
\def\prosima{$\; \buildrel \propto \over \sim \;$}
\def\gsim{\lower.5ex\hbox{\gtsima}}
\def\lsim{\lower.5ex\hbox{\ltsima}}
\def\simgt{\lower.5ex\hbox{\gtsima}}
\def\simlt{\lower.5ex\hbox{\ltsima}}
\def\simpr{\lower.5ex\hbox{\prosima}}

\def\ga{\gsim}

\title
[Clustering according to galaxy properties]
{The dependence of clustering on galaxy properties} 
\author[C. Li \etal]
{Cheng Li${^{1,2,3}}$ \thanks{E-mail: leech@ustc.edu.cn},
 Guinevere Kauffmann${^{3}}$, Y.P. Jing${^{1}}$, Simon D.M. White$^{3}$,
\newauthor Gerhard B\"orner$^{3}$, F.Z. Cheng$^{2}$\\
${^1}$The Partner Group of MPI f\"ur Astrophysik,
      Shanghai Astronomical Observatory,
      Nandan Road 80, Shanghai 200030, China \\
${^2}$Center for Astrophysics, University of Science
      and Technology of China, Hefei, Anhui 230026, China \\
${^3}$Max-Planck-Institut f\"ur Astrophysik,
      Karl-Schwarzschild-Strasse 1, 85748 Garching, Germany}

\date{
Accepted ........
Received .......;
in original form ......}

\pubyear{2006}
\begin{document}
\maketitle

\begin{abstract}
We use a sample of $\sim 200,000$ galaxies drawn from the Sloan Digital
Sky Survey (SDSS) with $0.01 < z <  0.3$ and $-23< M_{^{0.1}r} < -16$ to study
how clustering depends on properties such as stellar  mass ($M_\ast$),
colour ($g-r$), 4000\AA\ break strength (D$_{4000}$), concentration index
($C$), and stellar surface mass density ($\mu_\ast$).  Our measurements
of $w_p(r_p)$ as a function of r-band luminosity  are in excellent
agreement with previous two-degree Field Galaxy Redshift Survey and SDSS analyses. 
We compute $w_p(r_p)$ as
a function of stellar mass and we find that more massive galaxies cluster
more strongly than less massive galaxies, with the difference increasing
above the characteristic stellar mass $M^\ast$ of  the Schechter mass
function.  We then divide our sample according to colour, 4000\AA\ break
strength, concentration and surface density.  As expected, galaxies with
redder colours, larger 4000\AA\ break strengths, higher concentrations and
larger surface mass densities cluster more strongly. The clustering 
differences are largest on small scales and for low mass galaxies.
At fixed stellar mass, the dependences of clustering on colour and 4000
\AA\ break strength are  similar.  Different results are obtained when
galaxies are split by concentration or surface density.  The dependence
of $w_p(r_p)$ on $g-r$ and D$_{4000}$ extends out to physical scales that
are significantly larger than those of individual dark matter haloes
($> 5 h ^{-1} $ Mpc). This large-scale clustering dependence is not seen
for the parameters  $C$ or $\mu_\ast$. On small scales ($< 1 h^{-1}$ Mpc), 
the amplitude of the correlation function is constant for ``young''
galaxies with $1.1 < $D$_{4000} < 1.5$ and a steeply rising function of
age for  ``older'' galaxies with D$_{4000} > 1.5$.  In contrast,  the
dependence of the amplitude of $w_p(r_p)$ on concentration on scales less
than 1$h^{-1}$ Mpc is strongest for disk-dominated galaxies with $C<2.6$.
This demonstrates that different  processes are required to explain
environmental trends in the structure and in the star formation history of
galaxies.
\end{abstract}


\section{Introduction}
Our understanding of the large-scale structure of the Universe
has come primarily from studies of redshift surveys of nearby galaxies. 
The two-point correlation function (2PCF) of galaxies has long served
as the primary way of quantifying the clustering properties of 
galaxies in these surveys (for example, Peebles 1980).
As the fundamental lowest order statistic,
the 2PCF is simple to compute and provides a full statistical description
for Gaussian fields. It can also be easily
compared with the predictions of theoretical models.
Such comparisons have led to the conclusion that the observations are not
consistent with the predictions of the  standard $\Lambda$CDM ``concordance''
model unless there is a scale-dependent bias in the distribution of galaxies
relative to the dark matter 
(Jing, Mo \& B\"orner 1998; Jenkins \etal 1998; Gross \etal 1998).

Benson \etal (2000a) clarified how the dependence of galaxy formation
efficiency on halo mass could lead to just such a scale-dependent
bias. On large scales, the bias in the galaxy distribution
is related in a simple way to the bias in the distribution
of dark haloes. On small scales, the amplitude and slope
of the correlation function is determined by the interplay of a number
of different effects, including the distribution of
the number of galaxies that occupy a halo of given mass
and the fact that the  brightest galaxy in each halo is always located near
the halo centre.
These ideas have been further developed into the so-called ``halo occupation
distribution'' (HOD) approach  by many different authors
(for example Jing, Mo \& B\"orner 1998, Seljak 2000; 
Peacock \& Smith 2000; Berlind \& Weinberg  2002; Cooray \& Sheth 2002; 
Yang, Mo \& van den Bosch 2003).

The HOD approach enables one to understand why the correlation function
of $L^\ast$ galaxies is close to a power law over nearly
four orders of magnitude in amplitude in a flat, $\Omega_0=0.3$ CDM
universe. However, an important corollary is that the clustering
properties of galaxies ought to depend strongly on galaxy
colour, star formation rate and morphology, because
the halo occupation distributions of galaxies are predicted to depend
sensitively on these properties (see for example Kauffmann, Nusser
\& Steinmetz 1997; Kauffmann \etal 1999; Benson \etal 2000b).

The fact that the measured correlations of galaxies differ according
to type  has been known for almost three decades.
Davis \& Geller (1976)  computed angular correlations for
galaxies in the Uppsala Catalog and showed that elliptical-elliptical
correlations were characterized by a power law with steeper
slope than spiral-spiral correlations. Dressler (1980) quantified
this as a relation between galaxy type and local galaxy
density, with an increasing elliptical and S0 population 
and a corresponding decrease in spirals in the densest environments.

The large  redshift surveys assembled in recent years, e.g. 2dFGRS
and SDSS, have provided  angular positions and redshifts for
samples of hundreds of thousands of  galaxies and have
allowed the dependence of clustering on galaxy properties
to be studied with unprecedented accuracy.
These studies have established that the clustering of galaxies 
in the local Universe depends on a variety of factors, 
including luminosity
(Norberg \etal 2001, Zehavi \etal 2002, Zehavi \etal 2005),
colour (Zehavi \etal 2002, Zehavi \etal 2005), 
concentration (Zehavi \etal 2002, Goto \etal 2003),
and spectral type
(Norberg \etal 2002, Budav{\' a}ri \etal 2003, Madgwick \etal 2003).
These studies have revealed that galaxies with red colours, 
bulge-dominated morphologies and spectral types indicative of 
old stellar populations reside preferentially in dense regions
(Zehavi \etal 2005 and references therein, hereafter Z05).
Furthermore, luminous galaxies cluster more
strongly than less luminous galaxies, with the luminosity dependence
becoming more significant for galaxies brighter than $L^\ast$
(the characteristic luminosity of the Schechter [1976] function).
When galaxies are divided by colour,
redder galaxies show a higher amplitude
and steeper correlation function at all luminosities.

In order to interpret these clustering dependencies
in the framework of galaxy formation models, it is
useful to express the clustering results in terms of physical
quantities such as galaxy mass, size and mean stellar age, instead of
more traditional quantities such as luminosity or colour.
Galaxy luminosity does not necessarily  correlate very closely  
with stellar mass 
(the dominant baryonic component in all but the smallest galaxies). 
Both luminosity and colour are subject to
strong dependences on the fraction of young stars 
in the galaxy and on its dust content. These effects also
complicate comparisons between the clustering of
low redshift and high redshift galaxies.
It is now known that the star formation
rates in galaxies evolve very strongly as a function of redshift.
As a result, if one measures a change in clustering amplitude
at fixed luminosity, it is not simple to ascertain which part of the effect
is caused by the evolution in the stellar mass-to-light ratio ($M_\ast/L$)
and which part by a change in
the halo occupation distributions at higher redshift.                                

In this paper we study the dependence of
galaxy clustering on both luminosity and stellar mass using
a large sample of galaxies drawn from the Sloan Digital 
Sky Survey.
We then  probe the dependence on other physical 
parameters, including colour ($g-r$), 4000\AA\ break strength 
(D$_{4000}$), concentration parameter ($C$)and  
stellar surface mass density ($\mu_\ast$).
The first two quantities, i.e. $g-r$ and D$_{4000}$,
are parameters associated with the recent star formation history
of the galaxy (D$_{4000}$ is expected to be less sensitive to
 dust attenuation effects than colour),
whereas the other two are related to galaxy {\em structure}.
We first describe                         
the observational samples used for the analysis.
In \S3 we outline our method of measuring the 2PCF 
from large redshift surveys.
The results are described in \S4 and summarized
in the final section.

Throughout this paper,
We assume a cosmological model with the density parameter
$\Omega_0=0.3$ and the cosmological constant $\Lambda_0=0.7$.
To avoid the $-5\log_{10}h$ factor, the Hubble's constant
$h=1$, in units of $100\kms{\rm Mpc}^{-1}$,
is assumed throughout this paper when computing absolute magnitudes.
In this paper, the quantities with a superscript asterisk are those
at the characteristic luminosity/mass 
(e.g. characteristic luminosity $L^\ast$), 
whereas the quantities with a subscript asterisk refer to quantities
associated with the stars in a galaxy (e.g. stellar mass $M_\ast$).

\begin{figure}
\vspace{-3.2cm}
\centerline{\psfig{figure=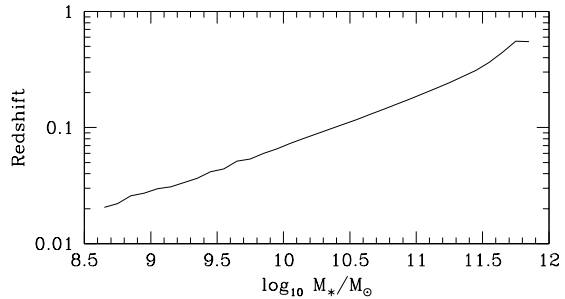,width=\hssize}}
\vspace{-0.2cm}
\caption{The results of
 a computation that we carried out to see out to what
 redshift a given mass in stars with maximum possible
 $M_\ast/L$ would be detected. This assumes a 13 Gyr
 stellar population and Bruzual \& Charlot models.}
\label{fig:range}
\end{figure}
\begin{figure}
\centerline{\psfig{figure=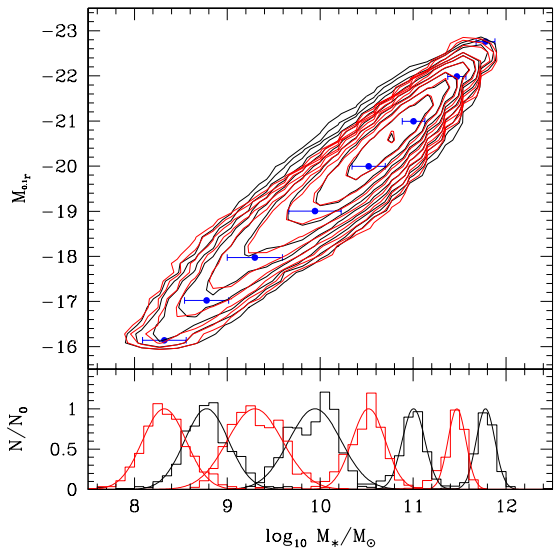,width=\hssize}}
\vspace{-0.2cm}
\caption{Shown in the {\it upper} panel are the contours of number density
of galaxies in the plane of stellar mass {\it vs} luminosity.
The {\it black} lines are for the data and the {\it red} are reconstructed
using the  Gaussian functions that best fit the stellar mass distribution
in 282 different luminosity intervals.
The contour levels are increased by factors of 2 from the lowest 
(15 $[0.2 mag]^{-1}[0.2\log_{10}M_\odot]^{-1}$) to
the highest (7680 $[0.2 mag]^{-1}[0.2\log_{10}M_\odot]^{-1}$).
The {\it lower} panel shows examples of the Gaussian distributions.
{\it Histograms} show  the data and  {\it solid} lines
the best-fits. $N_0$ is the Gaussian height.
The corresponding centers ({\it blue points}) and widths ({\it errorbars})
of the Gaussians are shown in the {\it upper} panel.}
\label{fig:mass_lum}
\end{figure}

\section{Observational Samples}

\subsection{NYU-VAGC}
The Sloan Digital Sky Survey (SDSS)
is the most ambitious optical imaging  and spectroscopic survey
to date. 
The survey goals are to obtain photometry of a quarter
of the sky and spectra of nearly one million objects.  Imaging is
obtained in the {\em u, g, r, i, z} bands (Fukugita \etal 1996;
Smith \etal 2002; Ivezi{\' c} \etal 2004) with a special purpose drift scan camera
(Gunn \etal 1998) mounted on the SDSS 2.5~meter telescope at
Apache Point Observatory.  The imaging data are photometrically
(Hogg \etal 2001) and astrometrically (Pier \etal 2003)
calibrated, and used to select stars, galaxies, and quasars for
follow-up fibre spectroscopy.  Spectroscopic fibres are assigned to
objects on the sky using an efficient tiling algorithm designed to
optimize completeness (Blanton \etal 2003b).  The details of the
survey strategy can be found in (York \etal 2000) and an
overview of the data pipelines and products is provided in the Early
Data Release paper (Stoughton \etal 2002).

The large areal coverage and moderately deep survey limit 
(a mean redshift of $\sim 0.1$ for galaxies in the main spectroscopic sample)
make the SDSS ideal for studying  large-scale structure
and the characteristics of  galaxy populations 
in the local Universe.
The SDSS covers two regions on the sky, one in the northern Galactic cap (NGC)
and another in the southern Galactic cap (SGC).
In the SGC, three stripes are observed, one along the celestial equator
and the other two north and south of the equator.
The NGC lies mostly above Galactic latitude 30$^\circ$,
but its footprint is adjusted slightly to lie within the minimum
of the Galactic extinction contours (Schlegel, Finkbeiner, \& Davis 1998),
resulting in an elliptical survey region (York \etal 2000).
Currently the survey in the NGC consists of two separate regions,
one  along the celestial equator (hereafter NGCE)
and another off the equator (hereafter NGCO).

In this paper we use the New York University Value Added Catalog
(NYU-VAGC)\footnote{http://wassup.physics.nyu.edu/vagc/},
which is a catalog of local galaxies (mostly below $z \approx 0.3$)
constructed by Blanton \etal (2005a) based on the SDSS Data Release Two
(DR2, Abazajian \etal 2004).
Earlier proprietary versions of this catalog have formed the basis
of many SDSS investigations of the power spectrum,
correlation function, and luminosity function of galaxies.
The current version of the NYU-VAGC consists of 693,319 photometric objects
(3514 deg$^2$); 343,568 of these have redshift determinations
(2627 deg$^2$), with about 85\% completeness.
This small subset of the full SDSS catalog
contains all of the information necessary
for analyzing the SDSS spectroscopic survey at the catalog level.
Compared with the catalogs distributed by the SDSS DR2 Archive Servers,
the NYU-VAGC is photometrically calibrated in a more consistent
way, reducing systematic calibration errors across the sky from
$\sim$ 2\% to about $\sim$ 1\%.
It is therefore more appropriate for statistical studies of galaxy peoperties,
galaxy clustering, and galaxy evolution.
The NYU-VAGC is described in detail in Blanton \etal (2005a).

\subsection{Physical quantities}

The rich stellar absorption-line spectrum of a typical SDSS galaxy provides 
unique information about its stellar content and  dynamics. 
Kauffmann \etal (2003a) presented a method for using this
information to estimate the stellar masses of galaxies. 
The amplitude of the 4000 \AA\ break (the narrow version of the index defined in
Balogh \etal 1999)
and the strength of the H$\delta$ absorption line (the Lick
$H\delta_A$ index of Worthey \& Ottaviani 1997) were used as diagnostics of the
stellar populations of the  galaxies. Both indices
were corrected for the observed contributions of the emission lines
in their bandpasses. From a library of 32,000 model star formation histories,
the measured D$_{4000}$ and $H\delta_A$ indices were used 
to obtain a maximum likelihood  estimate of  the  $z$-band $M_\ast/L$
for each galaxy. By comparing the colour predicted by the
best-fit model to the observed colour of the galaxy,
the attenuation of the starlight due to dust could be estimated.

The SDSS imaging data provide the basic structural parameters
 that are used in this analysis.
The $z$-band absolute magnitude, combined with the 
estimated values of $M_\ast/L$ and dust attenuation
$A_z$ yield the stellar mass ($M_\ast$). The half-light radius in the
$z$-band and the stellar mass yield the effective stellar surface mass-density
($\mu_\ast = M_\ast/2\pi r_{50,z}^2$, in unit of $h^2M_\odot/$kpc$^2$).
As a proxy for Hubble type we use
the SDSS ``concentration'' parameter $C$, which is defined as the ratio
of the radii enclosing 90\% and 50\% of the galaxy light in the $r$ band
(see Stoughton \etal 2002). Strateva \etal (2001) find that galaxies
with $C >$ 2.6 are mostly early-type galaxies, whereas spirals and irregulars
have 2.0 $< C <$ 2.6.

The reader is referred to Kauffmann \etal (2003a) for a more detailed
description of the methodology used to  derive  the stellar masses
used in this paper. An analysis of how the physical properties
of galaxies correlate with mass is presented in Kauffmann \etal (2003b).
All the parameters used  in this paper are available publically
at http://www.mpa-garching.mpg.de/SDSS/ (see also Brinchmann \etal 2004).

\subsection{Sample selection}\label{sec:sample}
\begin{table*}
\label{tbl:lum_samples}
\caption{Flux-limited samples selected according to luminosity/stellar mass}
\begin{center}
\begin{tabular}{clrrrrcrrrr} \hline\hline
&&\multicolumn{4}{c}{\sc Number of Galaxies}
&&\multicolumn{4}{c}{\sc Percentage in Subsamples$^a$}
\\ \cline{3-6} \cline{8-11}
{\sc Sample}&$M_{^{0.1}r}$&
{\sc SGC}&{\sc NGCO}&{\sc NGCE}&{\sc Total} & &$g-r$ &D$_{4000}$ &$C$ &$\log_{10}\mu_\ast$
\\ \hline
L1...........&$[-17.0,-16.0)$&458&548&608&1614
& &27.2\% &41.7\% &90.7\% & 48.0\%\\
L2...........&$[-17.5,-16.5)$&735&1261&1115&3111
& &26.2\% &32.2\% &82.9\% & 43.4\%\\
L3...........&$[-18.0,-17.0)$&1257&2301&1695&5253 
& &27.3\% &27.6\% &74.2\% & 41.4\%\\
L4...........&$[-18.5,-17.5)$&2130&3808&2693&8631
& &31.0\% &26.3\% &67.0\% & 44.2\%\\
L5...........&$[-19.0,-18.0)$&3657&6391&4366&14414
& &36.6\% &29.3\% &60.5\% & 49.3\%\\
L6...........&$[-19.5,-18.5)$&6532&10754&8582&25868
& &43.9\% &36.4\% &55.9\% & 56.3\%\\
L7...........&$[-20.0,-19.0)$&10349&16788&15740&42877
& &49.1\% &42.6\% &54.3\% & 61.5\%\\
L8...........&$[-20.5,-19.5)$&14804&24688&22879&62371
& &51.9\% &46.9\% &54.9\% & 63.7\%\\
L9...........&$[-21.0,-20.0)$&18460&31997&27530&77987
& &52.9\% &51.2\% &56.6\% & 61.9\%\\
L10..........&$[-21.5,-20.5)$&17717&31010&25376&74103
& &53.4\% &56.4\% &58.9\% & 54.7\%\\
L11..........&$[-22.0,-21.0)$&12140&20647&16252&49039
& &55.9\% &62.7\% &64.3\% & 41.1\%\\
L12..........&$[-22.5,-21.5)$&5384&8895&6876&21155
& &61.1\% &70.5\% &72.6\% & 23.3\%\\
L13..........&$[-23.0,-22.0)$&1267&2097&1674&5038
& &64.8\% &78.0\% &76.3\% & 7.20\%\\
\hline
{\sc Sample}&$\log_{10}M_\ast$
\\ \hline
M1...........&$[9.0,9.5)$&1686&3230&2325&7241
& &14.7\% &14.5\% &55.8\% &26.3\%\\
M2...........&$[9.5,10.0)$&4086&6695&5237&16018
& &23.8\% &19.8\% &44.3\% &36.5\%\\
M3...........&$[10.0,10.5)$&9757&15528&14275&39560
& &43.4\% &38.8\% &49.5\% &56.3\%\\
M4...........&$[10.5,11.0)$&17340&30519&26423&74282
& &55.3\% &53.1\% &58.9\% &63.8\%\\
M5...........&$[11.0,11.5)$&12475&21213&16671&50359
& &65.5\% &70.0\% &70.5\% &51.7\%\\
M6...........&$[11.5,12.0)$&1183&2082&1603&4868
& &71.7\% &83.7\% &78.1\% &17.3\%\\
\hline
\multicolumn{11}{l}{$^a$\ Percentage of objects in subsample with larger value
of physical quantities.}
\end{tabular}
\end{center}
\end{table*}
\begin{table*}
\label{tbl:vol_samples}
\caption{Volume-limited samples}
\begin{center}
\begin{tabular}{cllrrrr} \hline\hline
&&&\multicolumn{4}{c}{\sc Number of Galaxies}\\ \cline{4-7}
{\sc Sample}&$M_{^{0.1}r}$&
\multicolumn{1}{c}{$z$}&{\sc SGC}&{\sc NGCO}&{\sc NGCE}&{\sc Total}\\
\hline
VL1..........&$[-18.0,-17.0)$&$(0.01,0.03)$&675&1011&1016&2702\\
VL2..........&$[-19.0 -18.0)$&$(0.02,0.04)$&986&1776&1433&4195\\
VL3..........&$[-20.0,-19.0)$&$(0.03,0.07)$&4510&7135&4514&16159\\
VL4..........&$[-21.0,-20.0)$&$(0.04,0.07)$&2202&3275&2076&7553\\
VL5..........&$[-21.0,-20.0)$&$(0.04,0.10)$&6886&10021&10772&27679\\
VL6..........&$[-22.0,-21.0)$&$(0.07,0.16)$&5566&9446&8335&23347\\
VL7..........&$[-23.0,-22.0)$&$(0.10,0.23)$&705&1146&874&2725\\
\hline
{\sc Sample}&$\log_{10}M_\ast$&
\multicolumn{1}{c}{$z$}&{\sc SGC}&{\sc NGCO}&{\sc NGCE}&{\sc Total}\\
\hline
VM1..........&$[9.0,9.5)$  &$(0.015,0.045)$& 1195& 2190& 1624& 5009\\
VM2..........&$[9.5,10.0)$ &$(0.020,0.075)$& 3368& 5622& 4020&13010\\
VM3..........&$[10.0,10.5)$&$(0.025,0.100)$& 8039&12530&11865&32434\\
VM4..........&$[10.5,11.0)$&$(0.040,0.140)$&14324&24403&22035&60762\\
VM5..........&$[11.0,11.5)$&$(0.070,0.200)$&10343&17393&14065&41801\\
\hline
\end{tabular}
\end{center}
\end{table*}
In this paper,
all the three regions in NYU-VAGC, i.e. NGCE, NGCO and SGC, are considered.
Statistics are measured separately for the three regions
but the results are always presented for the whole survey
by combining the results in these regions.

We first select all NYU-VAGC galaxies
with extinction corrected Petrosian magnitude
$14.5 < r < 17.77$.
The bright limit  is so chosen because the SDSS becomes incomplete for bright
galaxies with large angular size,
whereas the faint limit corresponds to the magnitude limit of the Main galaxy sample
in SDSS.
Further criteria for galaxies to be included
in our analysis are, 1) they are identified as
galaxies from the Main sample
(see Blanton \etal 2005a for a detailed description),
2) they lie within the redshift range  $0.01 \le z \le 0.3$
and the absolute magnitude range $-23<M_{{0.1}_r}<-16$.
Here $M_{{0.1}_r}$ is the $r$-band absolute magnitude
corrected to its $z = 0.1$ value
using the $K-$correction code ({\tt kcorrect v3\_1b}) of Blanton \etal (2003a)
and the luminosity evolution model of Blanton \etal (2003c).
Our resulting sample includes a total of 196238 galaxies.

The galaxies are then divided into a variety of different subsamples.
We create 13 subsamples according to absolute magnitude, ranging from 
$M_{^{0.1}r}=-16$ to $M_{^{0.1}r}=-23$. Each sample includes galaxies 
in an absolute magnitude interval of 
1 magnitude, with successive subsamples overlapping by 0.5 magnitude.
Details are given in Table 1
(Samples L1-L13).

\begin{table}
\label{tbl:phy_samples}
\caption{Flux-limited samples selected according to physical quantities}
\begin{center}
\begin{tabular}{ccrrrr} \hline\hline
&&\multicolumn{4}{c}{\sc Number of Galaxies}\\ \cline{3-6}
{\sc Sample}&$g-r$&{\sc SGC}&{\sc NGCO}&{\sc NGCE}&{\sc Total}\\
\hline
c1...........&$[0.2,0.5)$&2348&3674&3383&9405\\
c2...........&$[0.3,0.6)$&5798&9511&8411&23720\\
c3...........&$[0.4,0.7)$&9350&15713&13647&38710\\
c4...........&$[0.5,0.8)$&11464&19502&16925&47891\\
c5...........&$[0.6,0.9)$&14206&23717&21363&59286\\
c6...........&$[0.7,1.0)$&16557&28504&25476&70537\\
c7...........&$[0.8,1.1)$&13188&22679&20229&56096\\
c8...........&$[0.9,1.2)$&6979&12617&10714&30310\\
\hline
{\sc Sample}&D$_{4000}$&{\sc SGC}&{\sc NGCO}&{\sc NGCE}&{\sc Total}\\
\hline
D1...........&$[1.0,1.3)$&3807&6245&5309&15361\\
D2...........&$[1.1,1.4)$&7755&12868&11050&31673\\
D3...........&$[1.2,1.5)$&10333&17565&15065&42963\\
D4...........&$[1.3,1.6)$&9900&17010&14561&41471\\
D5...........&$[1.4,1.7)$&8281&14296&12192&34769\\
D6...........&$[1.5,1.8)$&7762&13298&11527&32587\\
D7...........&$[1.6,1.9)$&9090&15716&13874&38680\\
D8...........&$[1.7,2.0)$&9902&17143&15622&42667\\
D9...........&$[1.8,2.1)$&8048&13776&12805&34629\\
D10..........&$[1.9,2.2)$&4274&7044&6916&18234\\
D11..........&$[2.0,2.3)$&1108&1654&1766&4528\\
\hline
{\sc Sample}&$C$&{\sc SGC}&{\sc NGCO}&{\sc NGCE}&{\sc Total}\\
\hline
C1...........&$[1.5,2.1)$&3372&5677&4783&13832\\
C2...........&$[1.7,2.3)$&7457&12720&10708&30885\\
C3...........&$[1.9,2.5)$&11027&19257&16399&46683\\
C4...........&$[2.1,2.7)$&12504&22439&19232&54175\\
C5...........&$[2.3,2.9)$&13007&23332&20616&56955\\
C6...........&$[2.5,3.1)$&12841&22119&19967&54927\\
C7...........&$[2.7,3.3)$&10620&17116&15881&43617\\
C8...........&$[2.9,3.5)$&6533&9827&9227&25587\\
C9...........&$[3.1,3.7)$&2647&3752&3580&9979\\
\hline
{\sc Sample}&$\log_{10}\mu_\ast$&{\sc SGC}&{\sc NGCO}&{\sc NGCE}&{\sc Total}\\
\hline
$\mu$1...........&$[8.00,8.50)$&2006&3415&2818&8239\\
$\mu$2...........&$[8.25,8.75)$&5430&9447&7989&22866\\
$\mu$3...........&$[8.50,9.00)$&9960&17419&15139&42518\\
$\mu$4...........&$[8.75,9.25)$&14166&24865&21974&61005\\
$\mu$5...........&$[9.00,9.50)$&13593&23073&20834&57500\\
$\mu$6...........&$[9.25,9.75)$&7015&10908&10044&27967\\
$\mu$7...........&$[9.50,10.0)$&1495&2065&1819&5379\\
\hline
\end{tabular}
\end{center}
\end{table}
Similarly, the galaxies are divided into 6 subsamples according
to  $\log_{10}M_\ast$ (M1-M6 in Table 1).
We do not consider galaxies  with $\log_{10}M_\ast<9$, because
the volume of the survey over which such systems can be 
detected is extremely small.
This is illustrated in 
Fig.\ref{fig:range}, where  we plot the maximum redshift out to which
a galaxy of mass $M_\ast$  with maximal $M_\ast/L$ would be detected in the survey.
This calculation  assumes a 13 Gyr  single-age 
stellar population and is based on the  Bruzual \& Charlot (2003) models.
At stellar masses below $10^9 M_\odot$, the oldest galaxies are only visible
at $z<0.03$. For galaxies with masses less than  $10^8 M_\odot$ the maximum redshift
is well below 0.02. 

To compare our results to previous work, we have also constructed
volume-limited subsamples (see Table 2), including subsamples
that are volume-limited in luminosity (Sample VL1-VL7) and in stellar mass
(Sample VM1-VM5).
The absolute magnitude ranges and redshift ranges used
for selecting  subsamples VL1-VL7  are the same as in Z05.

As will be described in Section 4.4, we further divide each
luminosity and stellar mass subsample into red and blue,
high D$_{4000}$ and low D$_{4000}$, low concentration
and high concentration, low density and high density subsamples by
fitting the distributions of these parameters using  bi-Gaussian functions.  
These subsamples are also listed in Table 1.
It is also interesting to investigate how  clustering varies as 
a function of colour/D$_{4000}$/concentration/surface density 
at fixed stellar mass.
To this end, we select a sample of galaxies
with stellar masses in the range of $10<\log_{10}M_\ast<11$,
and divide the galaxies into subsamples according to their
$g-r$ colours (Sample c1-c8), D$_{4000}$ values (Sample D1-D12), 
concentrations (Sample C1-C10) and surface mass densities (Sample $\mu$1-$\mu$6).
The details of these subsamples are given in Table 3.

\section{Clustering Measures}
In this section, we outline our  method for measuring the
galaxy two-point correlation function for a flux-limited 
sample of galaxies.                        
We begin by describing our methods for                      
constructing random samples.
We then describe how we  correct for the effect
of fibre collisions.
Finally, we describe the 2PCF estimator and how measurement errors are calculated.

\subsection{Constructing Random samples}\label{sec:random}
In order to use galaxy surveys in a statistically meaningful way,
we need to have complete knowlege of their selection effects.
A detailed account of the observational selection effects accompanies
the NYU-VAGC release.  The survey geometry is expressed
as a set of disjoint convex spherical polygons,
defined by a set of ``caps''. This methodology was developed
by Andrew Hamilton to deal accurately and efficiently 
with the complex angular masks of galaxy surveys
(Hamilton \& Tegmark 2002).
\footnote{http://casa.colorado.edu/$^\sim$ajsh/mangle/}
The advantage of using this method  is that
it is easy to determine whether a point is inside or outside
a given polygon (Tegmark, Hamilton \& Xu 2002).
The redshift sampling completeness
is then  defined as the number of galaxies with redshifts                      
divided by the total number of spectroscopic  targets in the polygon.
The completeness is thus a dimensionless number between 0 and 1,
and it is constant  within each of the polygons.
The limiting magnitude in each polygon is also provided (it changes slightly
across the survey region).

We have  constructed separate random catalogues for each of the
three regions of sky. These catalogues are designed to  include all observational
selection effects and are constructed as follows.
First, we select a spatial volume that is sufficiently large to 
contain the survey sample.
Then we randomly distribute points within the volume and eliminate
the points that are outside the survey boundary. Adopting the
same magnitude limits as in the observational sample, 
we select random galaxies and we use  the  luminosity function
derived by Blanton \etal (2003c) to assign to each of these galaxies an apparent
and  an absolute magnitude (appropriately $K$ and $E$-corrected, see \S\ref{sec:sample}).

Since we will estimate the correlation function as a function of
stellar mass, we also need to assign a mass  to each point in the random sample.
One way to do this is to use the observed  relation between luminosity and
stellar mass derived directly from our sample.
The black lines in the top panel of Fig.\ref{fig:mass_lum} 
show contours of the number density of galaxies in the  plane
of absolute magnitude {\it vs} stellar mass. It can be seen from the histograms
in the {\it bottom panel} of this figure that 
at fixed luminosity, 
the distribution of the stellar mass of galaxies is well described by a 
Gaussian, with the width of the Gaussian decreasing at higher luminosities.

We have divided the galaxies in our sample into 282 subsamples 
separated by 0.03 mag in $M_{^{0.1}r}$.
The bin size was chosen so that each subsample 
contained at least 500 galaxies.
The stellar mass distribution in each subsample is fitted with a Gaussian
and the solid lines in the  bottom panel of
Fig.\ref{fig:mass_lum}  show examples of these fits for 
several luminosity intervals.
To test the quality of the fits, 
we randomly assign each galaxy a {\it new} stellar mass  using the
Gaussian fits. The red lines in the top panel of 
Fig.\ref{fig:mass_lum}
show  contours of  the number density distribution that is  predicted by this
parametrization.
The recovered distribution is a good match to the observations 
except in the region
corresponding to luminous galaxies with low $M_\ast/L$s, where
the method tends to overpredict the masses.

\begin{figure*}
\vspace{-4.0cm}
\centerline{\psfig{figure=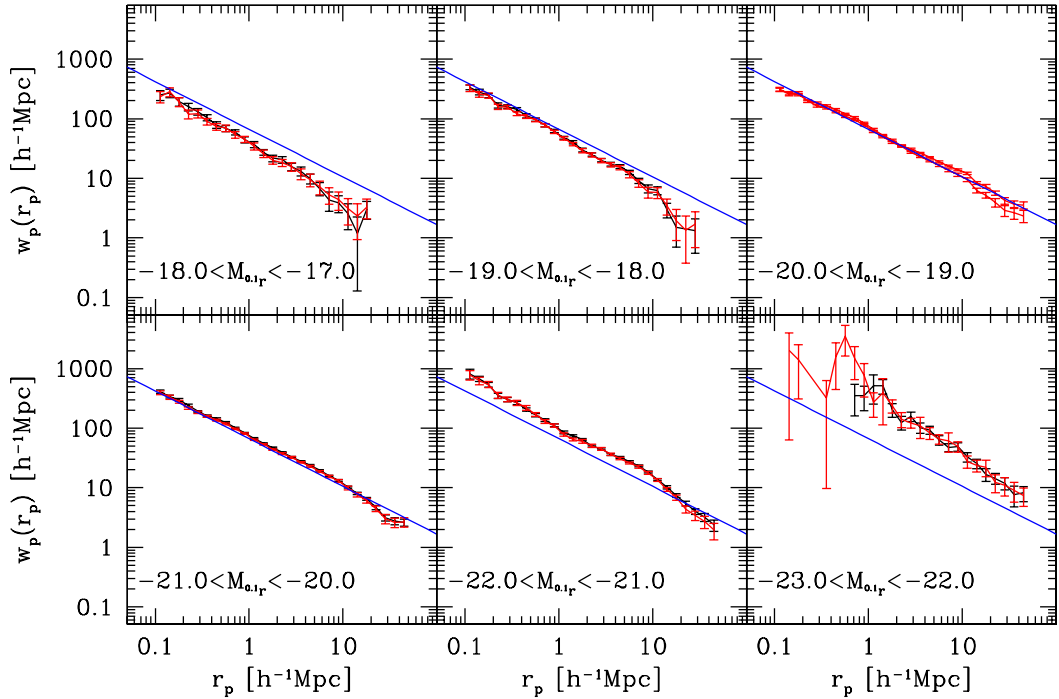,width=\hdsize}}
\vspace{-1.2cm}
\caption{Projected 2PCF $w_p(r_p)$ in different luminosity intervals 
(Sample L3, L5, L7, L9, L11 and L13 in Table 1).
When measuring 2PCFs, two methods are used to construct random
samples (see \S\ref{sec:random}).
The black lines are for the standard method and the red lines are for the
method in which the sky positions are randomized (see the
text for a detailed description).
In each panel, the blue line is the line
corresponding to $\xi(r)=(r/5 h^{-1} \mbox{Mpc})^{-1.8}$.
}
\label{fig:wrp_abm}
\end{figure*}

We now introduce a more general method, which should
still be applicable even when the relation between galaxy luminosity
and the physical property under investigation is not well
fit by a Gaussian and is subject to redshift-dependent
selection biases\footnote{One example of such a property
would be the  emission line luminosity of a central AGN.
The line detection limit is a strong function of redshift, because
increasing contamination by light from the surrounding host galaxy  makes extraction
of weak lines more difficult for more distant AGN.}.
Our method takes the observed sample and randomly re-assigns 
the position of each galaxy on the sky, while keeping the redshift,
absolute magnitude, stellar mass, and any other physical
quantities fixed. 
The spectroscopic incompleteness at each sky position
is imposed for the random points as in the observed sample.
To get a random catalogue as large as possible, we repeat the above
procedure for 20 times using different random number seeds.
In this way, all possible redshift-dependent selection biases 
are automatically taken into account, and it is only the sky position 
that is randomized. 
This method is valid only when the sample is a wide-angle
survey and the variation of its limiting magnitudes is small across the
survey region, both of which are valid in the SDSS.
For very large-area surveys such as the SDSS, randomizing the sky positions
should be sufficient to break the coherence of the large scale structures
in the survey.
In the next section we will  use random catalogues constructed using both
methods and we will show that the
measured projected correlation functions are in good agreement
(see Fig.\ref{fig:wrp_abm}).

\subsection{Volume Corrections}\label{sec:volume}
When computing correlation functions as a function of stellar mass,  
it is important to note that at a given stellar mass $M_\ast$,
galaxies with lower $M_\ast/L$ will be detected out to higher redshifts. 
A mass-selected sample will thus be biased to galaxies with younger
populations and this may lead to systematic errors
when computing the correlation function at fixed $M_\ast$.
In this paper, we correct for this $M_\ast/L$ bias 
by computing a weighted correlation function:
each galaxy pair is weighted by the inverse 
of the volume over which {\em both} galaxies can be detected 
in the survey.  This is similar to 
the $1/V_{max}$ correction that one makes when computing a mass 
function or luminosity function. 
The same volume weighting must also
be applied to the random catalogue. 
It is very simple to apply the same technique to the 
catalogues constructed by randomizing the sky positions, so
this will be our method of choice when estimating correlations
as a function of stellar mass.

In order to compute the volumes over which galaxies can be detected, we
have computed $z_{min}$ and  $z_{max}$ for each galaxy in the sample,
where $z_{min}$ is defined as the redshift where the galaxy has
an $r$-band magnitude of 14.5 and $z_{max}$ is the redshift where
the galaxy has an $r$-band magnitude of 17.77. These are derived using the
{\tt kcorrect} code of Blanton \etal (2003c).

\subsection{Correction for fibre collisions}
In the SDSS survey, two galaxies closer than $55^{\prime\prime}$
(corresponding to $\sim 100\kpch$ at the median redshift of our sample) 
cannot be assigned fibres simultaneously on one spectroscopic plate.
If these fibre ``collisions'' are  not taken into account,
the real-space (or projected) 2PCF 
will be systematically underestimated at small separations.
In earlier work (e.g. Zehavi \etal 2002, Tegmark \etal 2004),
a correction was made by simply assigning to each galaxy
affected by a  collision the same redshift as its
nearest spectroscopically-targeted  neighbour on the sky.
Zehavi \etal (2002) have performed extensive tests
of this procedure and have shown that it works well for $r_p>0.1\mpch$.
Tegmark \etal (2004) also find no evidence that fibre collisions
are boosting their measured power spectrum on the smallest
scales they probe ($k\sim0.3$\mpci).

In this paper, we use a different method for correcting for fibre collisions. 
We measure the angular 2PCF both for the spectroscopic samples and for
the parent photometric sample from which they were drawn; the effect
of fibre collisions can then be estimated and corrected for by
comparing the two correlation functions.
A similar method has been used in 2dF clustering analyses by 
Hawkins \etal (2003).

Here we briefly sumarize our method,
which will be described in more detail in a separate paper
(Li et al., in preparation).
We calculate the angular 2PCF for the photometric sample
($w_p(\theta)$) and for the spectroscopic sample ($w_z(\theta)$).
The quantity
\begin{equation}
F(\theta)=\frac{w_z(\theta)+1}{w_p(\theta)+1},
\end{equation}
can then be used to account for the effect of fibre collisions.
For each data-data pair,
we calculate the angular distance $\theta$ between the two members of
the pair and weight this pair by $1/F(\theta)$ when estimating the pair
counts.
If this correction is not applied, both $w(\theta)$ and $w_p(r_p)$
exhibit a strong ``rollover'' in amplitude on small scales.
Once the correction is applied, this feature disappears.
In the rest of our analysis, we will always include the                 
$1/F(\theta)$ weighting in the measurements of the correlation functions.
Since the effect of fiber collisions is expected to be
independent of galaxy property, we will not derive the correction
function $F(\theta)$ for each individual galaxy sample, but choose to
derive it from the whole sample and then apply it to our subsamples.

\subsection{Estimator of the Correlation Function and errors}
In this paper, the 2PCFs are measured in equal logarithmic 
bins of $r_p$ and in equal linear bins of $\pi$,
using the Hamilton (1993) estimator,
\begin{equation}
\xi(r_p,\pi) = \frac{4DD(r_p,\pi)RR(r_p,\pi)}{[DR(r_p,\pi)]^2}-1.
\end{equation}
Here $r_p$ and $\pi$ are the separations
perpendicular and parallel to the line of sight;
$DD(r_p,\pi)$ is the count of data-data pairs with
perpendicular separations in the bins $\log_{10}r_p \pm 0.5\Delta \log_{10}r_p$
and with radial separations in the bins $\pi \pm 0.5\Delta\pi$;
$RR(r_p,\pi)$ and $DR(r_p,\pi)$ are the counts of random-random
and data-random pairs, respectively.
The reason why we choose different bins                     
for $r_p$ and $\pi$ is the fact that $\xi(r_p,\pi)$
decreases rapidly as a function of  $r_p$, but remains
constant as a function of  $\pi$ on small scales.
Following standard practice, we estimate the projected two-point 
correlation function $w_p(r_p)$ by, 
\begin{equation}\label{eqwrpcal}
w_p(r_p) = 2\int_0^\infty\xi(r_p,\pi)d\pi 
= 2\sum_i\xi(r_p,\pi_i)\Delta\pi_i.
\end{equation}
Here the summation for computing $w_p(r_p)$ runs from $\pi_1 = 0.5 \mpch$
to $\pi_{40} = 39.5 \mpch$, with $\Delta\pi_i = 1 \mpch$.
The projected correlation function $w_p(r_p)$ is directly related to the real-space CF
$\xi(r)$ by a simple  Abel transform of $\xi(r)$.
Commonly, $w_p(r_p)$ is modelled by a power law
\begin{equation}
w\left(r_p\right) = Ar_p^{1-\gamma}.
\end{equation}
Then $\xi(r)$ is also a power law
\begin{equation}
\xi(r) = (r_0/r)^\gamma
\end{equation}
with
\begin{equation}
r_0^\gamma = \frac{A\Gamma(\gamma/2)}
{\Gamma(1/2)\Gamma[(\gamma-1)/2]},
\end{equation}
where $\Gamma(x)$ is the Gamma function.
However, the parametrization of the correlation function using only 
$r_0$ and $\gamma$ does not provide sufficient information to
recover the full observational results unless the correlation function 
is a pure power law on all scales. Our results (see below) 
show that this is not the case.
The departures of $w (r_p)$  from a pure power law have also
been discussed in previous papers (e.g. Zehavi \etal 2004).
We have thus chosen  to present our results in terms of 
the measured {\em amplitude} of $w_p(r_p)$
on different physical scales. We also tabulate 
the correlation functions so that our  readers can recover them accurately.
A detailed description of these tables (Tables 5 and 6) is given
in the Appendix. The tables themselves are available in electronic form
at http://www.mpa-garching.mpg.de/$^\sim$leech/papers/clustering/. 

The errors on the clustering measurements are estimated
using the bootstrap resampling technique 
(Barrow, Bhavsar, \& Sonoda 1984).
We generate 100 bootstrap samples from the observations
and compute the correlation functions
for each sample using the weighting scheme (but not the approximate
formula) given by Mo, Jing, \& B\"orner (1992).
The errors are then given by the scatter of the measurements among these
bootstrap samples.
The tests in Jing, Mo \& B\"orner (1998) using mock samples
showed that the bootstrap errors are comparable (within a factor of 2) to the
scatter among different mock samples, thus proving that the error 
estimates are robust.

\begin{figure}
\centerline{\psfig{figure=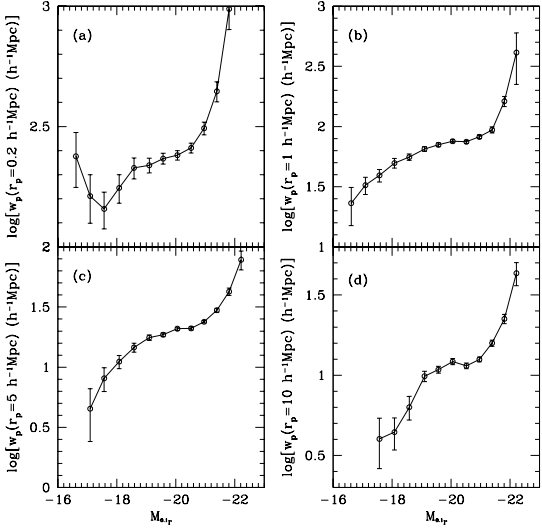,width=\hssize}}
\vspace{-0.2cm}
\caption{Amplitude of the projected 2PCF $w_p(r_p)$ as a function of luminosity
(Samples L1-L13) at $r_p=$0.2, 1, 5, and 10 $h^{-1}$ Mpc.}
\label{fig:wrp_abm_4panels}
\end{figure}
\begin{figure*}
\vspace{-4.0cm}
\centerline{\psfig{figure=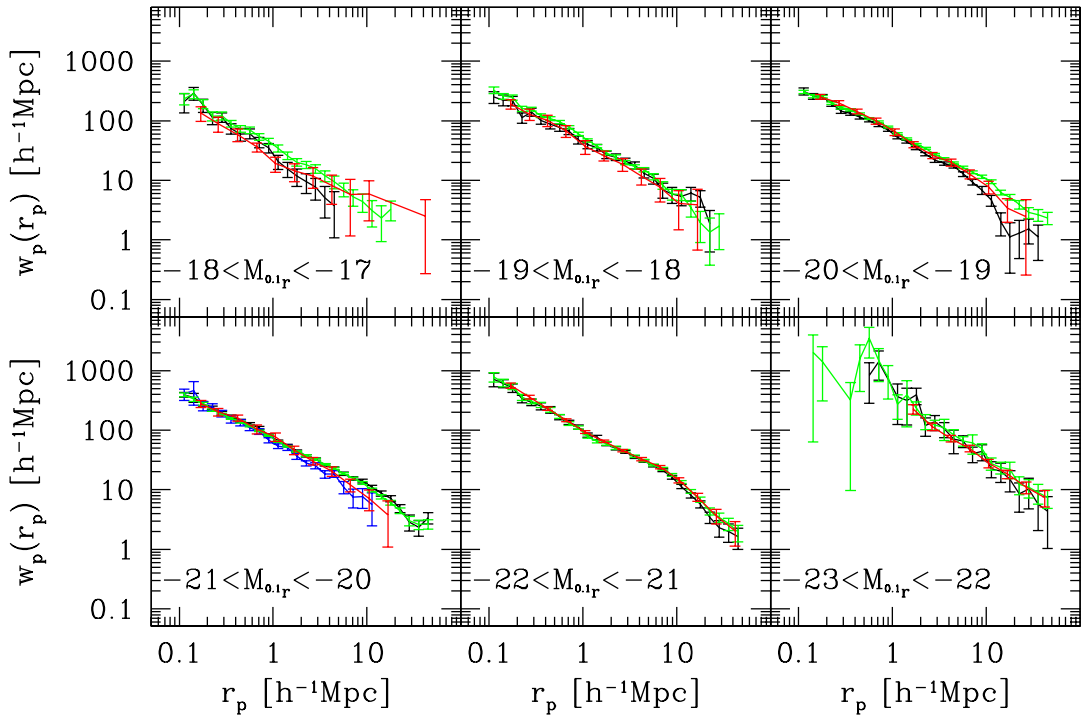,width=15cm}}
\vspace{-0.8cm}
\caption{Comparison of our  $w_p(r_p)$ measurments with
Zehavi \etal (2005, {\it red}). The {\it black} and {\it green}
are respectively for volume-limited and magnitude-limited samples.
The {\it blue} line in the left-bottom panel is for the volume-limited
sample with the redshift threshold reduced from 0.10 to 0.07. See
the text for a more  detailed description.
}
\label{fig:zehavi}
\end{figure*}
\begin{figure}
\vspace{-0.8cm}
\centerline{\psfig{figure=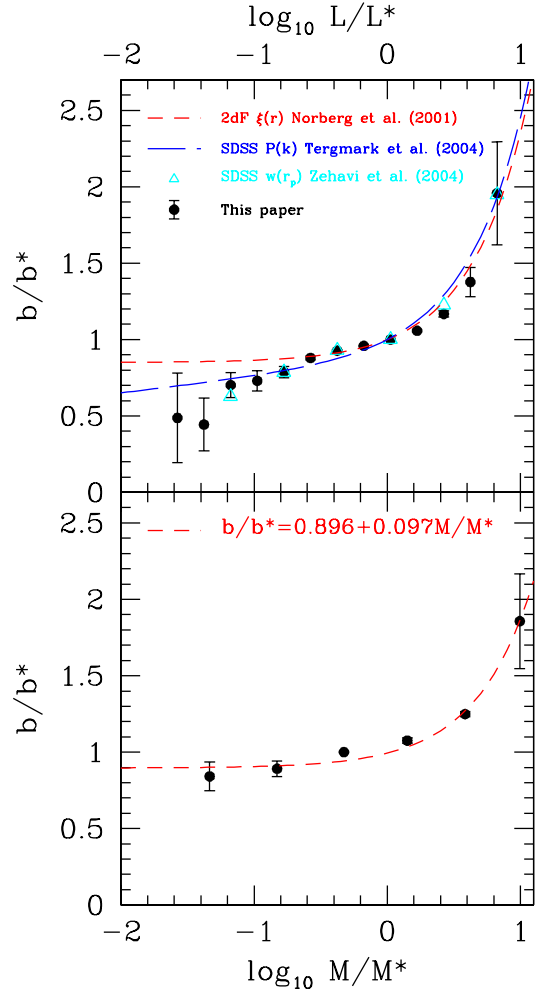,width=\hdsize}}
\vspace{-1.0cm}
\caption{
{\it Top panel:}
Relative bias factors for luminosity subsamples (Samples L1-L13).
Bias factors are defined by the relative amplitude of the $w_p(r_p)$
estimates at a fixed separation of $r_p=2.7\mpch$ and are normalized by the
$-21<M_{^{0.1}r}<-20$ sample (Sample L9, $L\approx L^\ast$).
The {\it dashed} curve is a fit to $w_p(r_p)$ measurments in the 2dF survey
$b/b^\ast=0.85+0.15L/L^\ast$ (Norberg \etal 2001), and the {\it long dashed}
curve is a fit obtained from measurements of the SDSS power spectrum,
$b/b^\ast=0.85+0.15L/L^\ast-0.04(M-M^\ast)$ (Tegmark \etal 2004; Note
that here the symboles $M$ and $M^\ast$ are for absolute magnitudes,
but not for stellar mass.).
The {\it triangles} are obtained from the $w_p(r_p)$ measurments
of Zehavi \etal (2005).
{\it Bottom panel:}
Relative bias factors for stellar mass subsamples (Sample M1-M6).
Bias factors are normalized by Sample M3, where  the mean stellar mass
is close to the characteristic stellar mass of the Schechter mass function.
}
\label{fig:bias}
\end{figure}

\section{Dependence of clustering on galaxy properties}

\subsection{Luminosity}

Fig.\ref{fig:wrp_abm} shows the projected 2PCF $w_p(r_p)$
in different luminosity intervals (Samples L3, L5, L7, L9, L11 and L13 in Table 1). 
The red and black lines on the figure compare the results obtained
for the two different methods of constructing random samples
described in Section 3.1. Black lines are for the ``standard'' method  
in which the selection function is explicitly modelled.
Red lines are for the method in which
the sky positions of the observed galaxies are randomly
re-assigned.
The agreement between the two methods is
very encouraging, suggesting that the latter method does
work well for analyzing large redshift surveys like the SDSS
and can be applied in the case of more complicated selection
by physical parameters with redshift-dependent biases.

To guide  the eye, we have plotted  the relation 
$\xi(r)=(r/5 h^{-1} \mbox{Mpc})^{-1.8}$
in blue in every panel in Fig.\ref{fig:wrp_abm}.
In general, we see that the amplitude of the correlation function
increases with luminosity, but the strength of this effect is 
different on different
scales. 
For galaxies fainter than $L^\ast$ ($M_{^{0.1}r}=-20.44$),
the clustering amplitude stays nearly constant 
on very small scales ($r_p\sim 0.1\mpch$), 
but  on larger scales there is a much stronger luminosity dependence.
For bright galaxies, the correlation amplitude increases strongly
with luminosity at all scales.
It is also interesting that the {\em slope} of the correlation function
gets flatter with increasing  luminosity for galaxies
fainter than $L^\ast$,  but then  increases for galaxies
brighter than $L^\ast$. In another word,  $L^\ast$ galaxies
exhibit the flattest correlation functions.

These trends are illustrated more clearly in 
Fig.\ref{fig:wrp_abm_4panels}, where we plot the amplitude of the
projected correlation function $w_p(r_p)$ as a function 
of luminosity  at $r_p=$0.2, 1, 5, and 10$\mpch$.
At $r_p$= 0.2 $\mpch$, the correlation function probes    
galaxy pairs that reside within a  common dark matter halo. 
 At $r_p$ = 10 $\mpch$
the correlation function should
only be sensitive to pairs of galaxies in separate haloes.   
This figure confirms that luminous galaxies
cluster more strongly than faint galaxies, with the difference
becoming more marked above $L^\ast$. However,
the luminosity dependence of galaxy clustering
is different on different scales.
On  small scales, the clustering amplitude does not 
vary with luminosity for galaxies fainter than $L^\ast$, but
increases steeply for galaxies brighter than $L^\ast$. In contrast, the amplitude
on large scales rises more continuously as a function of luminosity.
It is  also  interesting that the                           
dependence of $w_p(r_p)$ on luminosity appears to change slope at
$M_{^{0.1}r}\sim -20$. One possible reason for this switch
in behaviour  is that a significant 
fraction of faint galaxies are ``satellite'' systems           
orbiting within a common dark matter halo, whereas bright galaxies
are mainly ``central'' galaxies located at the centers of their dark matter haloes.
We intend to explore this in more detail in future work.

To compare our results to previous studies, we have also computed correlation
functions using samples that are volume-limited
in luminosity (Samples VL1-VL7). The results are shown
in Fig.\ref{fig:zehavi}.  Black lines show the correlation functions
for samples VL1-VL7. For comparison, the measurements
provided by Z05 are shown in red and  the correlation functions computed
from the corresponding magnitude-limited subsamples are shown in green.
The agreement between the magnitude-limited analysis and the volume-limited
one indicates that our results are  robust and reliable.
Furthermore, it can be seen that our measurements are 
in good agreement with those carried out by Z05, 
although there are some small differences.
These are probably due to the different 2PCF estimators or the different methods
of constructing random samples.
We note that the magnitude-limited sample of galaxies with  $-20<\absmag<-19$ 
(Sample L7) and the magnitude-limited and volume-limited samples
with  $-21<\absmag<-20$  (Sample L9 and VL5) all exhibit anomalously
high $w_p(r_p)$ values at large separations ($r_p\ga 5\mpch$).
As pointed out by Z05, this anomalous behavior is
a ``cosmic variance'' effect caused by an enormous supercluster
at $z\sim 0.08$, which overlaps  these three samples.
When the Sample VL5 is restricted to redshifts below 0.07,             
its projected correlation functions drops and steepens, 
({\it blue line} in Fig.\ref{fig:zehavi}),
coming into good agreement with that of Z05.

Following Z05, we calculate the relative bias factor $b/b^\ast$ as a function
of normalized luminosity $L/L^\ast$. The relative bias factor is defined
by the  amplitude of  $w_p(r_p)$ measured   at a fixed
separation $r_p = 2.7 h^{-1}$ Mpc relative to the value  measured for the
$-21<\absmag<-20$ subsample (Sample L9, which has $L\approx L^\ast$).
This fiducial separation of $2.7 h^{-1}$ Mpc was chosen because it is
well out of the very non-linear regime, but still small enough 
so that the correlation functions are very accurately measured in all surveys.
The solid circles in Fig.\ref{fig:bias} show our results and
the triangles show the SDSS results from Z05. The long dashed curve is taken from
Tegmark \etal (2004), where bias factors are derived from the galaxy
power spectrum $P(k)$ at wavelength $2\pi/k\sim 100\mpch$.
The results in this paper and in Z05, both of which are
derived from  $w_p(r_p)$ measurements, agree very well (as they should).
Our results are also in quite good agreement with those of Tegmark \etal 
The dashed curve in Fig.\ref{fig:bias} shows the result of
Norberg \etal (2001),
based on $w_p(r_p)$ measurements of somewhat more luminous  galaxies
in the 2dF survey ($\log_{10} L/L^\ast \ga -0.6$).
The agreement is again very good over the range of luminosities where 
the different analyses overlap.

\begin{figure*}
\vspace{-4.5cm}
\centerline{\psfig{figure=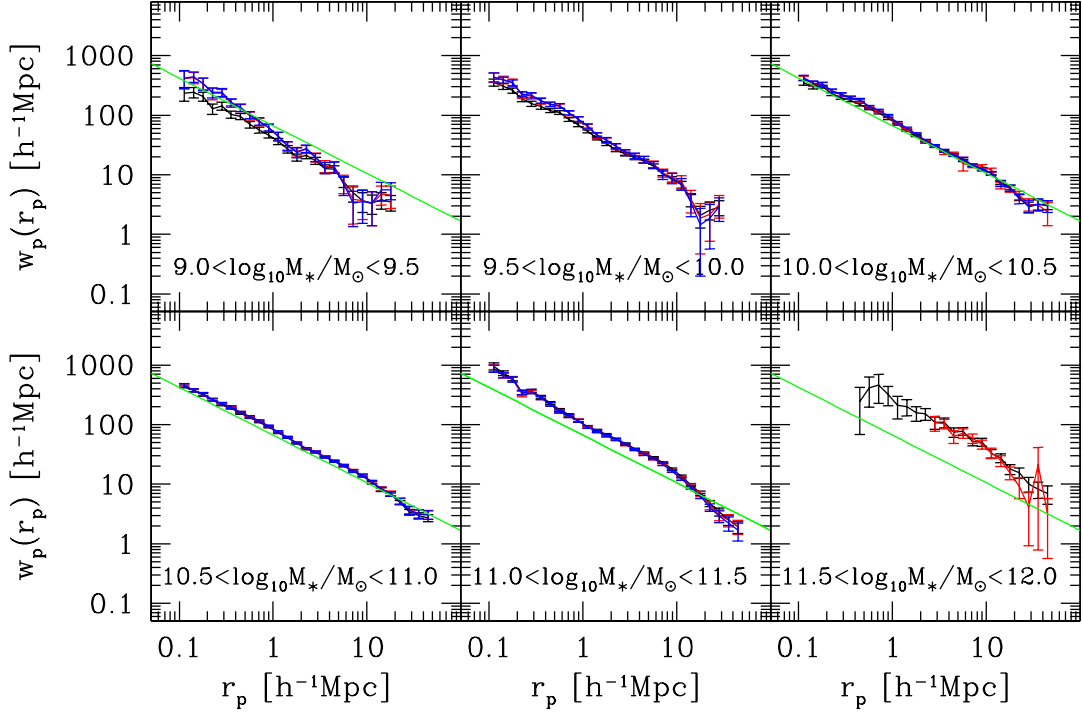,width=15cm}}
\vspace{-1.0cm}
\caption{Projected 2PCF for stellar mass subsamples, as indicated.
The {\it red} lines are for the results obtained by applying volume   
corrections, compared with those without applying the corrections
({\it black}). The {\it blue} lines in some panels are for the
samples that are volume-limited in $M_\ast$.
In each panel, the green line is the line
corresponding to $\xi(r)=(r/5 h^{-1} \mbox{Mpc})^{-1.8}$.
}
\label{fig:wrp_smass}
\end{figure*}
\begin{figure}
\vspace{-1.8cm}
\centerline{\psfig{figure=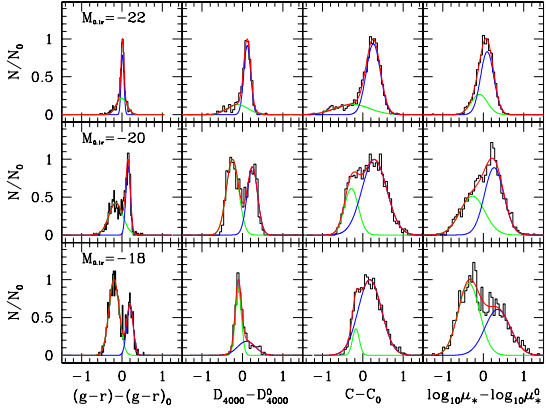,width=\hssize}}
\vspace{-0.2cm}
\caption{Examples of the bimodal distribution of physical quantities
in different luminosity intervals, as indicated.
In each panel, the {\it histogram} is
for the data, whileas the {\it green} and {\it blue} lines are the
best fit Gaussians and the {\it red} is the total.
$N_0$ is the maximum of the total fit, and the quantities with a zero
give  the median of the two Gaussian centers.}
\label{fig:bimodel}
\end{figure}
\begin{figure}
\vspace{-0.2cm}
\centerline{\psfig{figure=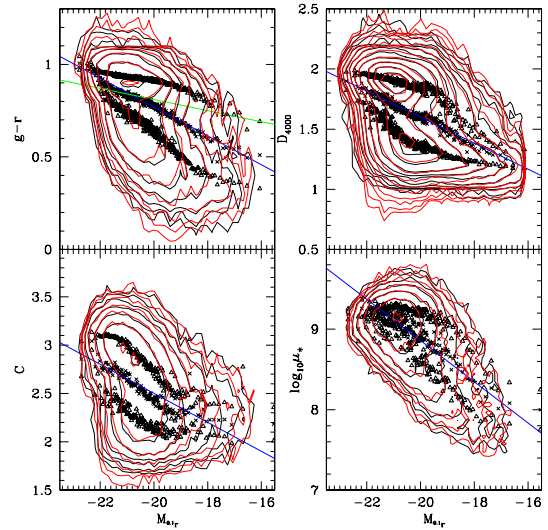,width=\hssize}}
\vspace{-0.3cm}
\caption{Contours of number density of galaxies
in the planes of luminosity {\it vs} physical quantities.
The {\it black} lines are for the data, whileas the
{\it red} are reconstructed according to the best-fitting
bi-Gaussians (see Fig.\ref{fig:bimodel}; also see the text
for a detailed description).
The {\it blue} lines are the best linear fits to the median Gaussian
centers as a function of luminosity
(see Fig.\ref{fig:bimodel}). These are  the luminosity-dependent cuts
that we adopt for dividing galaxies according to a given physical property.
The {\it green} line in the top-left panel is the $g-r$ cut adopted
by Zehavi \etal (2005).}
\label{fig:divider}
\end{figure}
\subsection{Stellar mass}

In this section, we present measurements of the projected
2PCF $w_p(r_p)$ as a function of stellar mass.
As discussed in section 3.2, when computing $w_p(r_p)$ as a 
function of mass, we weight each galaxy pair by the
inverse of the volume over which both galaxies can be detected
in the survey. The effect of this correction
can be seen in Fig.\ref{fig:wrp_smass} by comparing the black lines (no
volume-weighting) with the  red lines (with volume-weighting).
As can be seen, the volume correction steepens  the correlation
function of low mass galaxies. Recall that if no volume 
correction is applied, the sample is biased towards 
galaxies with low $M_\ast/L$ ratios. As we will show in detail in the
following section, the slope of the correlation function is
very sensitive to the colour (and hence the young stellar content)
of galaxies, particularly for low mass systems. This is why
the volume corrections make the most difference for
galaxies in our two lowest mass bins.
We have also compared our results with the measurements obtained using
samples that are volume-limited in stellar mass
({\it blue lines}). As can be seen, the results obtained for
the volume-limited samples agree very well  
with the volume-corrected $w_p(r_p)$. 

The bottom panel of Fig.\ref{fig:bias}  shows the relative bias factor
$b/b^\ast$ at $r_p=2.7\mpch$ as a function of stellar mass, with
points showing the results from our $w_p(r_p)$ measurements based on
samples M1-M6, and dashed lines showing the fit to the measurements
$b/b^\ast=0.90+0.10M/M^\ast$. The value $M^\ast$ is determined by
fitting a Schechter function to the stellar mass function of
the galaxies in our sample. We obtain 
$M^\ast=(4.11\pm0.02)\times 10^{10}h^{-2}M_\odot$,
$\alpha=-1.073\pm0.003$ and $\phi^\ast=0.0204\pm0.0001 h^3$Mpc$^{-3}$
(Wang \etal, in preparation).
Qualitatively, the behaviour of the relative bias as a function
of $M_\ast$ is very similar to the results obtained as a function
of $L$. This is not surprising, because luminosity
and stellar mass are reasonably tightly correlated (see Fig. 2).
What is of interest, however, is that these measurements can be used
to set constraints on the fraction of baryons that have been          
turned into stars in dark matter haloes of different mass.
We will come back to this in future work.

\subsection{Division by physical parameters}\label{sec:divider}

We now investigate how the clustering of galaxies of given
luminosity (or stellar mass) depends on properties such
as colour, 4000 \AA\ break strength, concentration and surface mass density.
Z05 performed such an analysis in the space of luminosity {\it vs} $g-r$ colour.
They adopted a tilted colour cut motivated by the colour-magnitude diagram.
A similar colour division is presented in Baldry \etal (2004),
who found that the distribution of galaxy colour could
be well approximate  using bi-Gaussian functions
(see Baldry \etal 2004, also see Fig.\ref{fig:bimodel} here).
Fig.\ref{fig:bimodel} shows that other physical
quantities, such as D$_{4000}$, $C$ and $\mu_\ast$ also
exhibit  bimodal distributions.
We thus fit bi-Gaussian functions to the distribution of
$g-r$, D$_{4000}$, $C$ and $\log \mu_\ast$ for each of
the  282 luminosity subsamples
described in \S\ref{sec:random}. These are shown in 
Fig.\ref{fig:bimodel} for three  representative luminosity intervals.
In Fig.\ref{fig:divider} we illustrate how well these fits recover
the true distribution of these parameters as a function of luminosity.
Black lines show contours of the actual number density of galaxies
and red lines show the predicted number densities from the
bi-Gaussian fits. As can be seen, the bi-Gaussian model
does a reasonable job of reproducing the observations.

\begin{table}
\label{tbl:divider}
\caption{Coefficients for the formula of dividers in physical quantities}
\begin{center}
\begin{tabular}{lrr}\hline\hline
Quantity   & \multicolumn{1}{c}{A} & \multicolumn{1}{c}{B} \\ \hline
$g-r$      & -0.788 $\pm$ 0.028 & -0.078 $\pm$ 0.001 \\
D$_{4000}$ & -0.563 $\pm$ 0.038 & -0.108 $\pm$ 0.002 \\
$C$        & -0.498 $\pm$ 0.193 & -0.150  $\pm$ 0.009  \\
$\log_{10}\mu_\ast$ & 3.738$\pm$0.213 & -0.256 $\pm$0.011\\
\hline
\end{tabular}
\end{center}
\end{table}
The division of the luminosity subsamples into red and blue, 
high D$_{4000}$ and low D$_{4000}$, high concentration
and low concentration, high surface density
and low surface density,  is defined as the mean of
the two Gaussian centers in each luminosity bin.
In Fig.~\ref{fig:divider}, triangles indicate the two Gaussian centers and
the crosses are the mean of these centers.
We fit the  dividing point as a function of luminosity  using a    
linear equation of the  form (see Fig.\ref{fig:divider}, {\it blue} lines),
\begin{equation}\label{eqn:divider}
P=A+B\cdot M_{^{0.1}r},
\end{equation}
where $P$ is the  physical parameter under
investigation, and $A$ and $B$ are the best-fitting
linear coefficients. These are listed in Table 4 for reference.
Using these best-fitting cuts (Eqn.\ref{eqn:divider}), 
we divide the galaxies in each of the 13 luminosity
samples (Sample L1-L13) and the 5 stellar mass samples (M1-M5) into
two further subsamples.
For simplicity, we use ``red" to denote the subsamples with larger
values of the physical quantity and ``blue" for the subsamples with 
the smaller value. The percentage of galaxies in the ``red"
subsamples are listed in the last 4 columns of Table 1.

\begin{figure*}
\vspace{-0.3cm}
\centerline{\psfig{figure=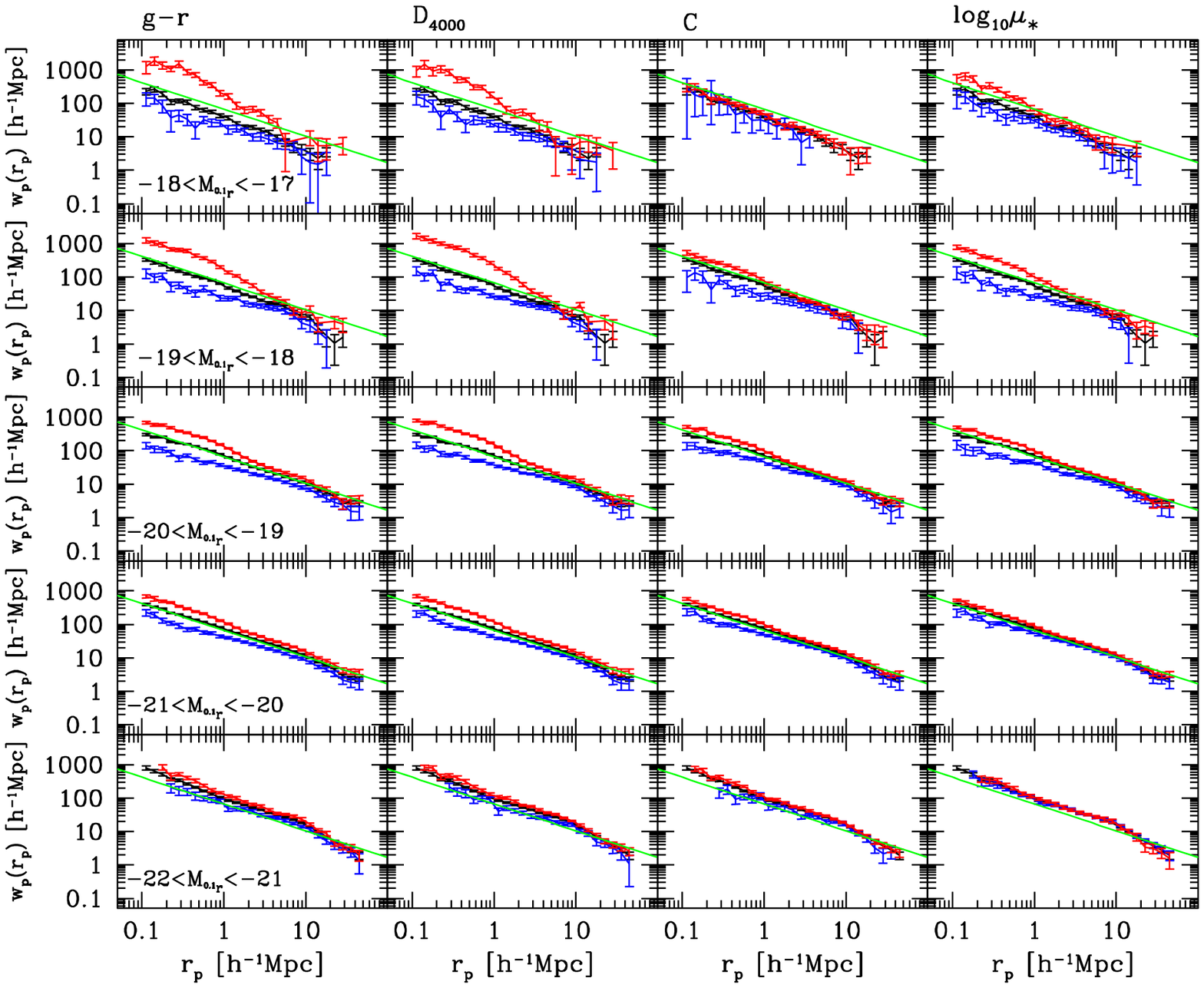,width=\hdsize}}
\vspace{-3.0cm}
\caption{Projected correlation function $w_p(r_p)$ for galaxies in different
luminosity intervals and with different properties
(from left to right: $g-r$ colour, D$_{4000}$, concentration
and $\log$ stellar surface mass density $\log_{10}\mu^\ast$).
The panels in each colume are for different luminosity subsamples,
with the range of absolute magnitude indicated in the left column.
In each panel, the {\it black} is for the full sample,
the {\it red} ({\it blue}) is for the subsample with
larger(smaller) value of the corresponding physical parameter.
In each panel, the green line is the line
corresponding to $\xi(r)=(r/5 h^{-1} \mbox{Mpc})^{-1.8}$.
}
\label{fig:wrp_par_abm}
\end{figure*}
\begin{figure*}
\vspace{-0.3cm}
\centerline{\psfig{figure=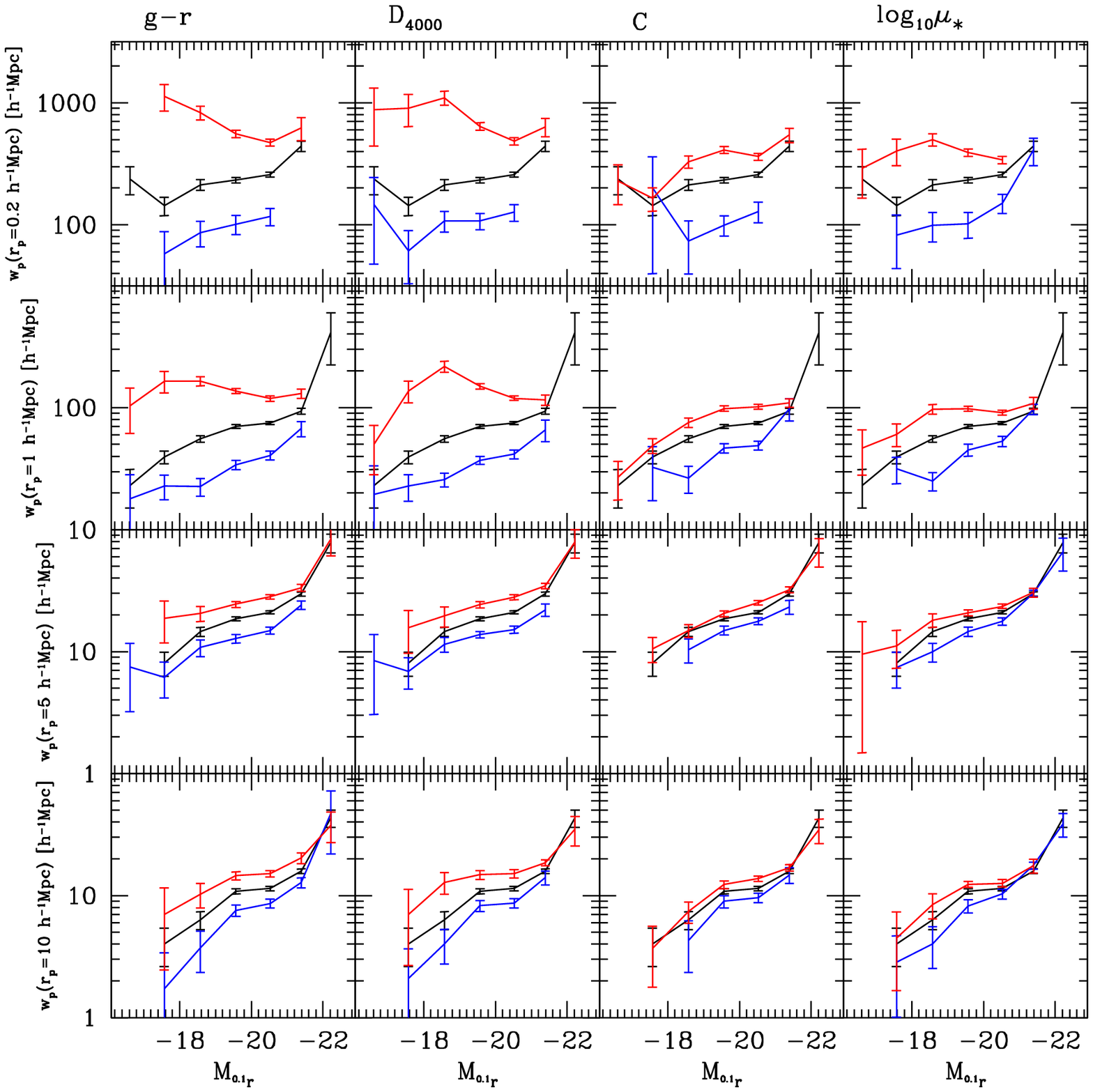,width=\hdsize}}
\vspace{-0.8cm}
\caption{$w_p(r_p)$ measured at $r_p=0.2,1,5,$ and $10\mpch$, as a function
of luminosity. The different columns show the dependence on different
physical quantities, as indicated. In each panel, the {\it black} is
for the full sample, the {\it red} ({\it blue}) is for the subsample with
larger(smaller) value of the corresponding physical parameter.}
\label{fig:amp_abm}
\end{figure*}
\begin{figure*}
\vspace{-0.3cm}
\centerline{\psfig{figure=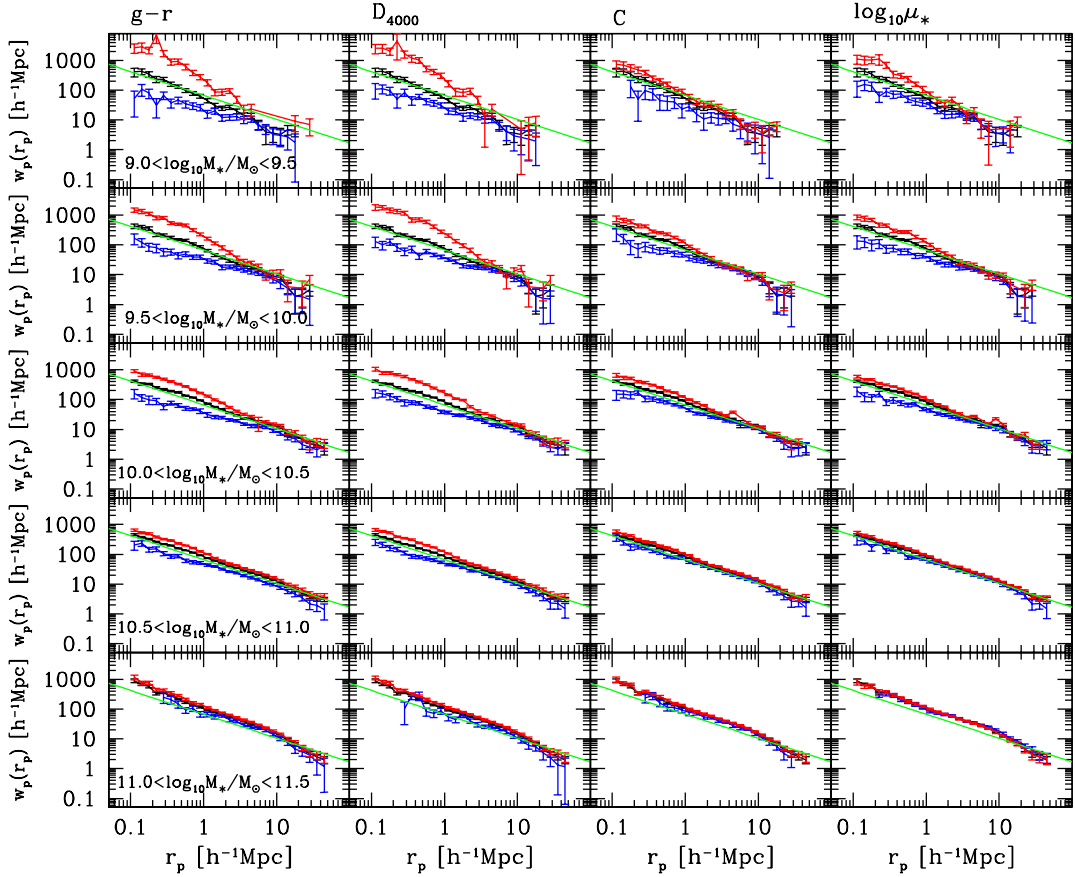,width=\hdsize}}
\vspace{-3.0cm}
\caption{Projected correlation function $w_p(r_p)$ for galaxies in
different stellar mass intervals and with different physical properties  
(from left to right: $g-r$ colour, D$_{4000}$, concentration
and $\log$ stellar surface mass density $\log_{10}\mu^\ast$).
The panels in each column are for different stellar mass subsamples,
with the range of stellar mass indicated in the left column.
In each panel, the {\it black} is for the full sample,
the {\it red} ({\it blue}) is for the subsample with
larger(smaller) value of the corresponding physical parameter,
and the green line is the line
corresponding to $\xi(r)=(r/5 h^{-1} \mbox{Mpc})^{-1.8}$.
}
\label{fig:wrp_par_smass}
\end{figure*}
\begin{figure*}
\centerline{\psfig{figure=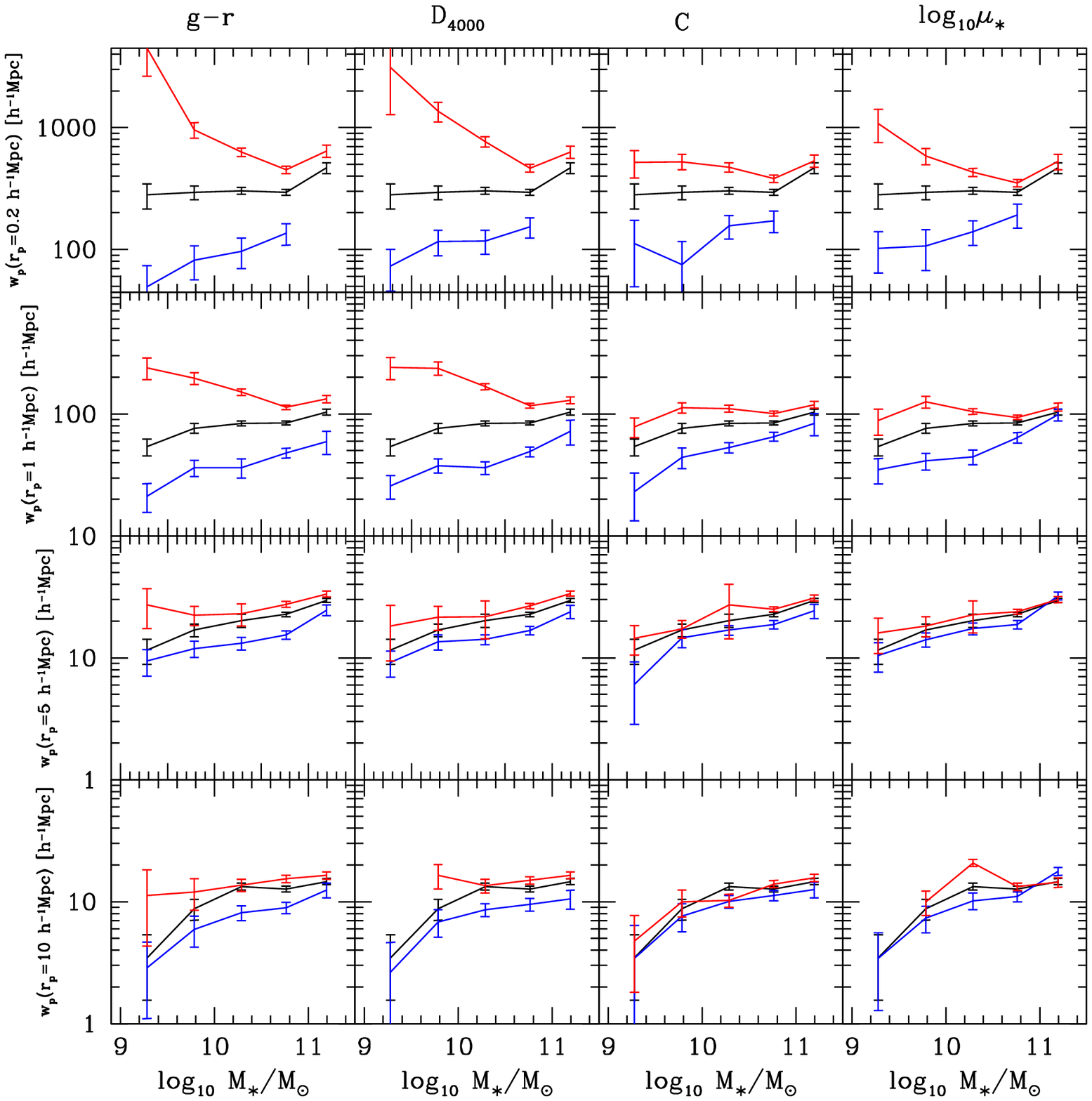,width=\hdsize}} 
\vspace{-0.8cm}
\caption{$w_p(r_p)$ measured at $r_p=0.2,1,5,$ and $10\mpch$, as a function 
of stellar mass. The different columns show the dependence on different  
physical quantities, as indicated. In each panel, the {\it black} is for the full sample, 
the {\it red} ({\it blue}) is for the subsample with
larger(smaller) value of the corresponding physical parameter.}
\label{fig:amp_smass}
\end{figure*}
\begin{figure*}
\centerline{\psfig{figure=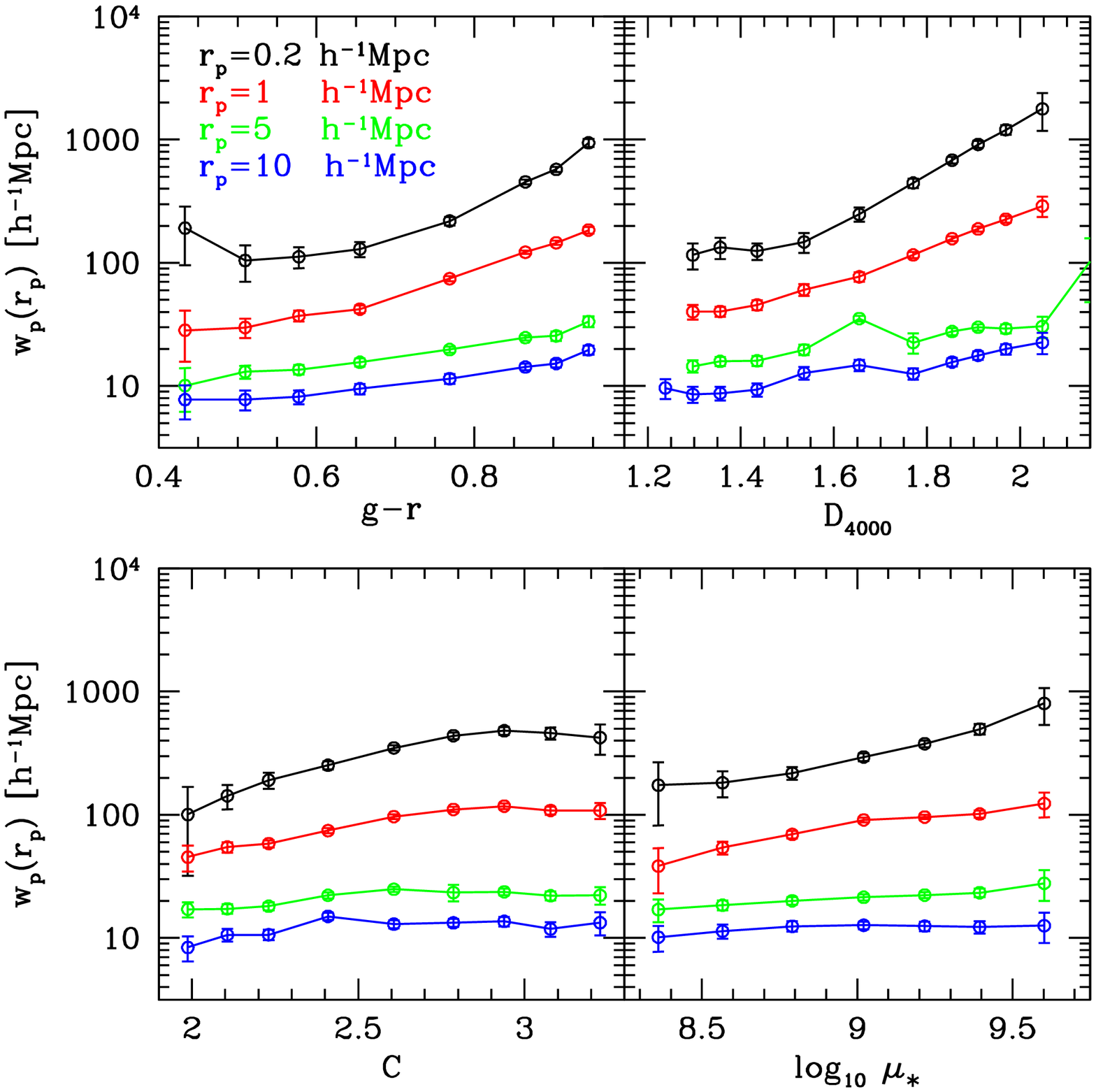,width=\hdsize}}
\vspace{-0.3cm}
\caption{Projected correlation functions measured at
$r_p=0.2,1,5$ and $10 h^{-1}$ Mpc, as a function of different
physical parameters.
In all panels, the stellar mass are limited to the
range  $10<\log_{10}M_\ast/M_\odot<11$.
}
\label{fig:amp_par}
\end{figure*}
\subsubsection{In luminosity bins}
The projected 2PCFs in the space of luminosity {\it vs}
colour, D$_{4000}$, concentration and surface density 
are presented  in Fig.\ref{fig:wrp_par_abm}.   Red (blue) lines
correspond to the  ``red" (``blue") subsamples.
Black lines are for the sample as a whole.                               
Fig.~\ref{fig:amp_abm} shows the measurements of the amplitude
of $w_p(r_p)$  at $r_p=$
0.2, 1, 5 and 10 $\mpch$.

When the sample is divided by $g-r$ colour,
redder galaxies of all luminosities  are more strongly clustered 
and have steeper correlation functions than their blue counterparts.
This  colour dependence
is much stronger for faint galaxies than for bright 
galaxies, particularly on small scales.
Fig.~\ref{fig:amp_abm} shows that the clustering amplitude    
of blue galaxies increases as a function of luminosity
at all scales.  However, the situation is more complicated
for red galaxies. On small scales, faint red galaxies are
clustered more strongly than bright red galaxies. 
On large scales, however,  the trend reverses and the clustering amplitude
increases with luminosity.
These results are all consistent with the findings of Z05.
The behaviour of the slope of the correlation function 
as a function of luminosity is also different for red and
blue galaxies. The correlation function of faint red
galaxies is very steep and the slope flattens systematically
as luminosity increases. In contrast, the slope of the
correlation function of blue galaxies exhibits rather
little change with luminosity. All these trends are qualitatively consistent
with a picture in which faint red galaxies are primarily
``satellite'' systems in massive dark matter haloes, but
faint blue galaxies occupy haloes of smaller mass 
(Z05; Berlind \etal 2005; Li \etal 2006).

Our results show that the dependence of clustering on D$_{4000}$ 
is very similar to what is obtained for $g-r$ colour.
On the other hand, rather different results are obtained for the  structural
parameters $C$ and $\mu_\ast$. 
Fig.~\ref{fig:amp_abm} clearly shows  that 
the dependence of $w_p(r_p)$ on  $g-r$/D$_{4000}$ is
considerably stronger than the dependence on  $C$/$\mu_\ast$ at
all physical scales.

\subsubsection{In stellar mass bins}
The projected 2PCFs in the space of stellar mass {\it vs} the same set       
of physical parameters
are presented  in Fig.\ref{fig:wrp_par_smass}.
The measurements at $r_p=$ 0.2, 1, 5 and 10$\mpch$ are plotted in
Fig.\ref{fig:amp_smass}.

Qualitatively, the results shown in Figs.~\ref{fig:amp_abm} and                             
\ref{fig:amp_smass} appear very similar.                                     
However, careful comparison of these two figures    
shows that  interesting quantitative differences
do exist between the clustering of the                
``red'' and ``blue'' subsamples at fixed luminosity and at fixed stellar mass.
On small scales, the dependences are stronger when evaluated at fixed
mass, particularly for low mass galaxies. We also note that
there is  a small difference in the clustering
amplitude of the ``red'' and ``blue'' subsamples at  projected
radii as large as 10 $h^{-1} $ Mpc in Fig.~\ref{fig:amp_abm}.
This difference is seen both in $g-r$ colour and D$_{4000}$
and more weakly in the structural parameters $C$ and $\mu_\ast$. 
Fig.  \ref{fig:amp_smass} shows, however, that at fixed stellar mass 
there is no longer any significant  difference in the clustering amplitude of
high concentration and low concentration galaxies
or high surface density and low surface density galaxies on scales
larger than 5 $h^{-1}$ Mpc. The clustering differences
in $g-r$ and D$_{4000}$ do persist, however.
This is a rather surprising result, because at scales larger than a few Mpc,
galaxies inside the same dark matter halo no longer contribute
to the clustering signal. Our result thus indicates
that at fixed stellar mass,  
the clustering
properties of the surrounding  {\em dark matter haloes} are
somehow  correlated with the colour of the selected galaxies.

To investigate this effect further, we have computed the 2PCF as a function    
of $g-r$, D$_{4000}$, $C$ and $\mu_\ast$ for galaxies spanning
a narrow range in stellar mass ($10^{10}-10^{11} M_{\odot}$).
In Fig.\ref{fig:amp_par}, we plot the amplitude of the correlation
function as a function of these quantities measured on
four different physical scales  
($r_p$= 0.2,1, 5 and 10$\mpch$).  
This figure confirms that the dependence of 
$w_p(r_p)$ on $g-r$/D$_{4000}$ extends 
out to larger physical scales than the 
dependence of $w_p(r_p)$ on $C$/$\mu_\ast$. 
The figure also shows that the dependence of $w_p(r_p)$
on $C$ and $\mu_\ast$  is also qualitatively quite different on
small scales. 
On  scales less than  $< 1 h^{-1}$ Mpc, 
the amplitude of the correlation function is 
constant for ``young'' galaxies with $1.1 < $D$_{4000} < 1.5$ and a steeply rising
function of age for  ``older'' galaxies with D$_{4000} > 1.5$. 
In contrast, the dependence
of the amplitude of  $w_p(r_p)$ on concentration is strongest for disk-dominated
galaxies with $C<2.6$ on these same scales. 
This demonstrates that different physical processes are required to explain
environmental trends in
star formation and in galaxy structure.

\section{Summary and Discussion}

In this paper we present our determinations of
the projected two-point correlation function (2PCF) $w_p(r_p)$ 
for different classes of galaxies in order to study
the dependence of clustering on the
physical properties of these systems. We use  
the New York University Value Added Catalog (NYU-VAGC) 
which is constructed from the   
the Sloan Digital Sky Survey Data Release Two (SDSS DR2).

The conclusions of this paper can be summarized as follows:
\begin{enumerate}
\item We confirm previous findings that luminous galaxies                       
cluster more strongly than faint galaxies, 
with the difference becoming larger  for galaxies 
with $L>L^\ast$, where $L^\ast$ is the characteristic luminosity
of the Schechter (1976) function. 
The dependence of galaxy clustering on luminosity  
is different on different physical scales.
On  small scales ($r_p\sim 0.2\mpch$), 
the correlation amplitude  is almost constant                        
for galaxies fainter than $L^\ast$, but the amplitude
increases sharply above $L^\ast$.
On large scales, the correlation amplitude increases more
continuously  as a  function of luminosity.       
Around $L^\ast$ there appears to be a shoulder,
with $w_p(r_p)$ increasing more steeply with $L$ for higher-luminosity
galaxies.
Our results are in good agreement with previous studies
of clustering as a function of luminosity in the SDSS.

\item We present 
$w_p(r_p)$ as a function of stellar mass.
In analogy with previous results
obtained as a function of luminosity,  
we find that more massive galaxies cluster more strongly than less massive
galaxies, with the difference increasing above the 
characteristic stellar mass $M^\ast$ of  the Schechter mass function. 

\item When galaxies are divided according to their physical
properties, we find that galaxies with                              
redder colours, larger 4000\AA\ break strengths, 
more concentrated structure, and higher surface mass densities
cluster more strongly and have steeper correlation functions
at all luminosities and masses.  The
differences in clustering strength are larger  on small scales 
and for low-luminosity and less massive  galaxies.

\item 
We have found that the dependence of $w_p(r_p)$ on $g-r$ or D$_{4000}$
extends out to larger physical scales ($r_p>5\mpch$) than the 
dependence of $w_p(r_p)$ on $C$ or $\mu_\ast$.
On  small scales ($\sim 0.2\mpch$), 
the behaviour of $w_p(r_p)$ as a function of $g-r$ or D$_{4000}$ 
and as a function of $C$ are qualitatively different. 

\end{enumerate}

We have chosen not to express our results in terms
of  power-law fits to our  $w_p(r_p)$
measurements. 
We have tabulated the measurements
of our correlation functions so that they can be accurately recovered. 
As discussed by Z05, a single power-law is a poor description
of the data and as we expand our exploration of physical parameter
space, it is important not to place unnecessary restrictions
on the way in which the observational results are described.
In this paper, we have chosen to  plot trends  in clustering {\em amplitude}
evaluated on a variety different physical scales. This leads to  a number of 
interesting insights that have not received
much attention up to now : (1) the dependence
of the clustering amplitude (or equivalently, the relative
bias factor) on luminosity is qualitatively different
on small scales and on large scales,  (2) there is a  different scale
dependence in the amplitude of the correlation
function  for parameters that measure the 
star formation histories of galaxies and  for parameters that measure
galaxy structure, suggesting that 
the trends in star formation
and in galaxy structure are governed  by different physical
processes.

Finally, it is worth comparing our results with the many studies that have
examined correlations between galaxy properties and the local environment.
One of the most fundamental correlations between the properties
of galaxies in the local Universe is the so-called
morphology-density relation. 
Oemler (1974) and Dressler (1980) pioneered the quantification
of this relation, showing that spheroidal systems 
reside preferentially in dense regions.
Since the standard morphological classification scheme mixes elements
that depend on the structure of a galaxy with elements related to
its recent star formation history, it is by no means obvious that these
two elements should depend on environment in the same way.

Recent studies using large surveys such as the SDSS  
have revealed that galaxy colour is the galaxy
property most predictive of the local environment
(e.g. Blanton \etal 2005b; Kauffmann \etal 2004).
Hogg et al. (2003) show that the local density increases strongly
with luminosity for the brightest galaxies. For faint galaxies, 
local density is senstive mainly to color, with faint
red galaxies occupying highest-density regions.
Blanton \etal (2005b) found that at fixed luminosity and colour, 
density is not closely related to surface brightness or to 
the S$\acute{e}$rsic index (a quantity related to galaxy structure),
so that morphological properties of galaxies are less closely related
to galaxy environment than their luminosities  and star formation histories.
Kauffmann \etal (2004) obtained very similar results.
They found that at fixed stellar mass both star formation and
nuclear activity depend strongly on local density, while
structural parameters such as size and concentration are almost
independent of it. 

Our analyses of $w_p(r_p)$ as a function of luminosity, colour
and structural parameters are consistent with these conclusions.
The power of the $w_p(r_p)$ statistic is that it encapsulates
information about how  galaxy properties depend on
environment over a wide range  of physical scales. 
Kauffmann \etal (2004) found no evidence for a significant
dependence of galaxy structure on local density.
However, their local densities are calculated in a fixed aperture
of 2 $\mpch$, whereas our plots (see Fig.\ref{fig:amp_par}) show
clearly that the dependence of  structural parameters  on environment becomes
significant  on scales that are smaller than this value.

The other advantage of $w_p(r_p)$ is that it is can be very easily compared
with the predictions of  galaxy formation simulations.
It probes the  physical processes occurring inside individual
dark matter haloes as well the masses of the dark matter haloes that host galaxies
of given mass, luminosity, size, age and concentration, thus placing strong
constraints  on theoretical models.
This will be the focus of  future work.

\section*{Acknowledgments}

We are grateful to Dr. Idit Zehavi for providing her $w_p(r_p)$
measurements and for her detailed comments on our paper,
and to Dr. Michael Blanton for his help
with the NYU-VAGC. We thank the SDSS teams for making
their data publicly available.
This work is supported by NKBRSF(G19990754), by NSFC(Nos.10125314, 10373012, 10073009),
by Shanghai Key Projects in Basic research (04jc14079, 05xd14019),
by the Max Planck Society,
and partly by the Excellent Young Teachers Program of MOE, P.R.C.
CL acknowledges the financial support of the exchange program between
Chinese Academy of Sciences and the Max Planck Society.

Funding for the creation and distribution of the 
SDSS Archive has been provided by the Alfred P. Sloan Foundation, 
the Participating Institutions, the National Aeronautics and Space Administration, 
the National Science Foundation, the U.S. Department of Energy, 
the Japanese Monbukagakusho, and the Max Planck Society. 
The SDSS Web site is http://www.sdss.org/.
The SDSS is managed by the Astrophysical Research Consortium (ARC) 
for the Participating Institutions. The Participating Institutions 
are The University of Chicago, Fermilab, the Institute for Advanced Study, 
the Japan Participation Group, The Johns Hopkins University, 
the Korean Scientist Group, Los Alamos National Laboratory, 
the Max-Planck-Institute for Astronomy (MPIA), the Max-Planck-Institute 
for Astrophysics (MPA), New Mexico State University, 
University of Pittsburgh, University of Portsmouth, 
Princeton University, the United States Naval Observatory, 
and the University of Washington.

\setcounter{table}{4}
\begin{table*}
\label{tbl:wrp_abm1}
\caption{$w_p(r_p)$ for galaxies in different luminosity intervals and with different properties}
\begin{center}
\begin{tabular}{crrrrrrrrr} \hline\hline
\multicolumn{1}{c}{$r_p$} & 
\multicolumn{1}{c}{All}   &
\multicolumn{2}{c}{$g-r$} &
\multicolumn{2}{c}{D$_{4000}$} &
\multicolumn{2}{c}{$C$} &
\multicolumn{2}{c}{$\log_{10}\mu_\ast$}
\\ \cline{3-4}\cline{5-6}\cline{7-8}\cline{9-10}
\multicolumn{1}{c}{[$h^{-1}$Mpc]} &&
\multicolumn{1}{c}{red}  &
\multicolumn{1}{c}{blue} &
\multicolumn{1}{c}{red}  &
\multicolumn{1}{c}{blue} &
\multicolumn{1}{c}{red}  &
\multicolumn{1}{c}{blue} &
\multicolumn{1}{c}{red}  &
\multicolumn{1}{c}{blue}
\\ \hline
\multicolumn{10}{l}{$-17.0\le M_{^{0.1}r}<-16.0$} \\
        0.11 &  319.6/111.5&  
 321.6/172.1 &       ------& 
1075.4/439.2 &  184.8/153.1& 
1912.7/1330.0&   94.7/114.3&  
 710.8/377.0 &  200.2/147.6\\
        0.14 &  221.0/62.9 &  
 235.5/79.2  &       ------&  
 800.2/319.6 &  127.8/104.8& 
1859.3/706.8 &   32.4/52.2 &  
 547.8/212.1 &    9.7/84.2 \\
        0.18 &  247.2/64.9 &
 246.5/102.1 &       ------&
 446.5/191.9 &  158.1/130.1&
 845.8/382.8 &  112.5/76.0 &
 365.1/142.4 &  180.7/150.8\\
        0.23 &  204.7/56.4 &
 180.1/58.4  &       ------&
1430.1/734.4 &  105.2/60.6 &
1014.5/409.5 &  363.2/608.3&
 244.1/104.5 & 1461.9/1130.4\\
\multicolumn{10}{l}{......} \\
\hline
\end{tabular}
\end{center}
\end{table*}
%



\appendix

\section[]{Description of Online Tables}

Tables 5 and 6 contain our $w_p(r_p)$ measurements for the galaxies with different
luminosities/stellar masses. Results are also
given as a function of the  physical quantities $g-r$, D$_{4000}$,
$C$ and $\mu_*$.
The tables are available in electronic form at
{\em http://www.mpa-garching.mpg.de/$^\sim$leech/papers/clustering/}.

Tables 5 and 6 list the data points in Figs.\ref{fig:wrp_par_abm} and
\ref{fig:wrp_par_smass} respectively, i.e. the measured
projected 2PCF $w_p(r_p)$ for galaxies in different luminosity or stellar mass
intervals and with different properties. Table 5 consists of 13 separate parts
corresponding to the 13 luminosity samples (Samples L1-L13 in Table 1).
Likewise, Table 6 consists of 5 parts corresponding to the 5 stellar mass
samples (Samples M1-M5 in Table 1). To make the description clearer,
we present here an abridged version for the first part in Table 5, listing only
the first several rows for Sample L1.
The first column is the projected separation $r_p$ in unit of $\mpch$, ranging
from $\sim 0.1\mpch$ up to $\sim 45\mpch$. The other columns give  the $w_p(r_p)$ 
measurements and errors for the full luminosity/stellar mass sample 
(Column 2) and the "red" and "blue" subsamples divided by $g-r$ 
(Columns 3-4), D$_{4000}$ (Columns 5-6), $C$ (Columns 7-8) and 
$\log_{10}\mu_\ast$ (Columns 9-10). 
Short straight lines  denote the points that have no measurements, 
either because of low S/N or for  any other reason.


\begin{thebibliography}{99}

\bibitem[Abazajian \etal (2004)]{abazajian04} Abazajian, K., \etal \ 2004, \aj, 128, 502 

\bibitem[\protect\citeauthoryear{Baldry et al.}{2004}]{2004ApJ...600..681B} Baldry I.~K., Glazebrook K., Brinkmann J., Ivezi{\' c} {\v Z}., Lupton R.~H., Nichol R.~C., Szalay A.~S., 2004, ApJ, 600, 681 

\bibitem[\protect\citeauthoryear{Balogh \etal }{1999}]{1999ApJ...527...54B} Balogh M.~L., Morris S.~L., Yee H.~K.~C., Carlberg R.~G., Ellingson E., 1999, ApJ, 527, 54 

\bibitem[Barrow, Bhavsar, \& Sonoda(1984)]{bbs84} Barrow, J.~D., Bhavsar, S.~P., \& Sonoda, D.~H.\ 1984, \mnras, 210, 19 

\bibitem[\protect\citeauthoryear{Benson \etal }{2000b}]{2000MNRAS.316..107B} Benson A.~J., Baugh C.~M., Cole S., Frenk C.~S., Lacey C.~G., 2000b, MNRAS, 316, 107  
 
\bibitem[\protect\citeauthoryear{Benson \etal }{2000a}]{2000MNRAS.311..793B} Benson A.~J., Cole S., Frenk C.~S., Baugh C.~M., Lacey C.~G., 2000a, MNRAS, 311, 793  

\bibitem[\protect\citeauthoryear{Berlind \& Weinberg}{2002}]{2002ApJ...575..587B} Berlind A.~A., Weinberg D.~H., 2002, ApJ, 575, 587 

\bibitem[\protect\citeauthoryear{Berlind et al.}{2005}]{2005ApJ...629..625B} Berlind A.~A., Blanton M.~R., Hogg D.~W., Weinberg D.~H., Dav{\'e} R., Eisenstein D.~J., Katz N., 2005, ApJ, 629, 625 

\bibitem[Blanton \etal (2003a)]{blanton03a} Blanton, M.~R., Brinkmann, J., Csabai, I., Doi, M., Eisenstein, D.~J., Fukugita, M., Gunn, J.~E., Hogg, D.~W. \& Schlegel, D.~J.\ 2003a, \aj, 125, 2348  

\bibitem[Blanton \etal (2003b)]{blanton03b} Blanton, M.~R., Lin, H., Lupton, R.~H., Maley, F.~M., Young, N., Zehavi, I., \& Loveday, J.\ 2003b, \aj, 125, 2276 

\bibitem[Blanton \etal (2003c)]{blanton03c} Blanton, M.~R., Lin, H., Lupton, R.~H., Maley, F.~M., Young, N., Zehavi, I., \& Loveday, J.\ 2003c, \apj, 592, 819 

\bibitem[\protect\citeauthoryear{Blanton \etal }{2005a}]{2005AJ....129.2562B} Blanton M.~R., \etal, 2005a, AJ, 129, 2562 

\bibitem[\protect\citeauthoryear{Blanton \etal }{2005b}]{2005ApJ...629..143B} Blanton M.~R., Eisenstein D., Hogg D.~W., Schlegel D.~J., Brinkmann J., 2005b, ApJ, 629, 143

\bibitem[\protect\citeauthoryear{Brinchmann \etal }{2004}]{2004MNRAS.351.1151B} Brinchmann J., Charlot S., White S.~D.~M., Tremonti C., Kauffmann G., Heckman T., Brinkmann J., 2004, MNRAS, 351, 1151 

\bibitem[\protect\citeauthoryear{Bruzual \& Charlot}{2003}]{2003MNRAS.344.1000B} Bruzual G., Charlot S., 2003, MNRAS, 344, 1000 

\bibitem[\protect\citeauthoryear{Budav{\' a}ri \etal }{2003}]{2003ApJ...595...59B} Budav{\' a}ri T., \etal, 2003, ApJ, 595, 59 

\bibitem[\protect\citeauthoryear{Cooray \& Sheth}{2002}]{2002PhR...372....1C} Cooray A., Sheth R., 2002, PhR, 372, 1 

\bibitem[\protect\citeauthoryear{Davis \& Geller}{1976}]{1976ApJ...208...13D} Davis M., Geller M.~J., 1976, ApJ, 208, 13  

\bibitem[Dressler(1980)]{dressler80} Dressler, A.\ 1980, \apj, 236, 351 

\bibitem[\protect\citeauthoryear{Fukugita \etal }{1996}]{1996AJ....111.1748F} Fukugita M., Ichikawa T., Gunn J.~E., Doi M., Shimasaku K., Schneider D.~P., 1996, AJ, 111, 1748 

\bibitem[Goto \etal (2003)]{goto03} Goto T., Yamaguchi C., Fujita Y., Okamura S., Sekiguchi M., Smail I., Bernardi M., Gomez P., 2003, \mnras, 346, 601

\bibitem[\protect\citeauthoryear{Gross \etal }{1998}]{1998MNRAS.301...81G} Gross M.~A.~K., Somerville R.~S., Primack J.~R., Holtzman J., Klypin A., 1998, MNRAS, 301, 81 

\bibitem[\protect\citeauthoryear{Gunn \etal }{1998}]{1998AJ....116.3040G} Gunn J.~E., \etal, 1998, AJ, 116, 3040 

\bibitem[Hamilton(1993)]{hamilton93} Hamilton, A.~J.~S.\ 1993, \apj, 417, 19 

\bibitem[Hamilton \& Tegmark(2002)]{ht02} Hamilton, A.~J.~S.~\& Tegmark, M.\ 2002, \mnras, 330, 506 

\bibitem[Hawkins \etal (2003)]{hawkins03} Hawkins, E., \etal \ 2003, \mnras, 346, 78  

\bibitem[\protect\citeauthoryear{Hogg \etal }{2001}]{2001AJ....122.2129H} Hogg D.~W., Finkbeiner D.~P., Schlegel D.~J., Gunn J.~E., 2001, AJ, 122, 2129 

\bibitem[\protect\citeauthoryear{Hogg et al.}{2003}]{2003ApJ...585L...5H} Hogg D.~W., et al., 2003, ApJ, 585, L5 

\bibitem[\protect\citeauthoryear{Ivezi{\' c} \etal }{2004}]{2004AN....325..583I} Ivezi{\' c} {\v Z}., \etal, 2004, AN, 325, 583 

\bibitem[\protect\citeauthoryear{Jenkins \etal }{1998}]{1998ApJ...499...20J} Jenkins A., \etal, 1998, ApJ, 499, 20 

\bibitem[Jing \& B{\" o}rner(2004)]{jb04} Jing, Y.~P. \& B{\" o}rner, G.\ 2004, \apj, 617, 782 

\bibitem[Jing, Mo, \& B{\" o}rner(1998)]{jmb98} Jing, Y.~P., Mo, H.~J. \& B{\" o}rner, G.\ 1998, \apj, 494, 1

\bibitem[\protect\citeauthoryear{Kauffmann, Nusser, \& Steinmetz}{1997}]{1997MNRAS.286..795K} Kauffmann G., Nusser A., Steinmetz M., 1997, MNRAS, 286, 795 

\bibitem[\protect\citeauthoryear{Kauffmann \etal }{1999}]{1999MNRAS.303..188K} Kauffmann G., Colberg J.~M., Diaferio A., White S.~D.~M., 1999, MNRAS, 303, 188 

\bibitem[\protect\citeauthoryear{Kauffmann \etal }{2003a}]{2003MNRAS.341...33K} Kauffmann G., \etal, 2003a, MNRAS, 341, 33 

\bibitem[\protect\citeauthoryear{Kauffmann \etal }{2003b}]{2003MNRAS.341...54K} Kauffmann G., \etal, 2003b, MNRAS, 341, 54 

\bibitem[Kauffmann \etal (2004)]{kauffmann04} Kauffmann, G., \etal \ 2004, \mnras, 353, 713 

\bibitem[\protect\citeauthoryear{Li et al.}{2006}]{2006MNRAS.368...37L} Li C., Jing Y.~P., Kauffmann G., Boerner G., White S.~D.~M., Cheng F.~Z., 2006, MNRAS, 368, 37

\bibitem[\protect\citeauthoryear{Madgwick \etal }{2003}]{2003MNRAS.344..847M} Madgwick D.~S., \etal, 2003, MNRAS, 344, 847 

\bibitem[Miller \etal (2003)]{miller03} Miller C.~J., Nichol R.~C., Gomez P.~L., Hopkins A.~M., Bernardi M., 2003, \apj, 597, 142

\bibitem[Mo, Jing, \& B\"orner(1992)]{mjb92} Mo, H.~J., Jing, Y.~P., B\"orner, G.\ 1992, \apj, 392, 452 

\bibitem[Norberg \etal (2001)]{norberg01} Norberg, P., \etal \ 2001, \mnras, 328, 64  

\bibitem[Norberg \etal (2002)]{norberg02} Norberg, P., \etal \ 2002, \mnras, 332, 827  

\bibitem[Oemler(1974)]{oemler74} Oemler, A.\ 1974, \apj, 194, 1

\bibitem[\protect\citeauthoryear{Peacock \& Smith}{2000}]{2000MNRAS.318.1144P} Peacock J.~A., Smith R.~E., 2000, MNRAS, 318, 1144 

\bibitem[Peebles(1980)]{peebles80} Peebles, P.J.E.\ 1980, The Large-Scale Structure of the Universe, Princeton University Press, Princeton 

\bibitem[\protect\citeauthoryear{Pier \etal }{2003}]{2003AJ....125.1559P} Pier J.~R., Munn J.~A., Hindsley R.~B., Hennessy G.~S., Kent S.~M., Lupton R.~H., Ivezi{\' c} {\v Z}., 2003, AJ, 125, 1559 

\bibitem[\protect\citeauthoryear{Schechter}{1976}]{1976ApJ...203..297S} Schechter P., 1976, ApJ, 203, 297 

\bibitem[Schlegel, Finkbeiner, \& Davis(1998)]{sfd98} Schlegel, D.~J., Finkbeiner, D.~P, \& Davis, M.\ 1998, \apj, 500, 525 

\bibitem[\protect\citeauthoryear{Seljak}{2000}]{2000MNRAS.318..203S} Seljak U., 2000, MNRAS, 318, 203 

\bibitem[\protect\citeauthoryear{Smith \etal }{2002}]{2002AJ....123.2121S} Smith J.~A., \etal, 2002, AJ, 123, 2121 

\bibitem[\protect\citeauthoryear{Stoughton \etal }{2002}]{2002AJ....123..485S} Stoughton C., \etal, 2002, AJ, 123, 485 

\bibitem[\protect\citeauthoryear{Strateva \etal }{2001}]{2001AJ....122.1861S} Strateva I., \etal, 2001, AJ, 122, 1861 

\bibitem[Tegmark, Hamilton \& Xu(2002)]{thx02} Tegmark, M., Hamilton, A.~J.~S., \& Xu,Y.\ 2002, \mnras, 335, 887

\bibitem[Tegmark \etal (2004)]{tegmark04} Tegmark, M., \etal \ 2004, \apj, 606, 702 

\bibitem[\protect\citeauthoryear{Worthey \& Ottaviani}{1997}]{1997ApJS..111..377W} Worthey G., Ottaviani D.~L., 1997, ApJS, 111, 377 

\bibitem[\protect\citeauthoryear{Yang, Mo, \& van den Bosch}{2003}]{2003MNRAS.339.1057Y} Yang X., Mo H.~J., van den Bosch F.~C., 2003, MNRAS, 339, 1057 

\bibitem[York \etal (2000)]{york00} York, D.~G., \etal \ 2000, \aj, 120, 1579 

\bibitem[Zehavi \etal (2002)]{zehavi02} Zehavi, I., \etal \ 2002, \apj, 571, 172 

\bibitem[Zehavi \etal (2004)]{zehavi04b} Zehavi, I., \etal \ 2004, \apj, 608, 16 

\bibitem[\protect\citeauthoryear{Zehavi et al.}{2005}]{2005ApJ...630....1Z} Zehavi I., et al., 2005, ApJ, 630, 1 

\end{thebibliography}
\end {document}